\documentclass[aps,prd,preprintnumbers,groupedaddress,nofootinbib,amssymb,notitlepage,eqsecnum]{revtex4-2}
\usepackage{here}
\usepackage[dvipdfmx]{graphicx}
\usepackage{amsmath,amsthm,amssymb}
\usepackage{bm}
\usepackage{color}
\usepackage{comment}

\usepackage{amsfonts}
\usepackage{dcolumn}
\usepackage{hyperref}
\allowdisplaybreaks[1]
\usepackage{stackengine}

\newcommand{\be}{\begin{equation}}  
\newcommand{\ee}{\end{equation}}
\newcommand{\ba}{\begin{eqnarray}}
\newcommand{\ea}{\end{eqnarray}}

\newcommand{\rd}{{\rm d}}

\newcommand{\bem}{\begin{bmatrix}}
\newcommand{\eem}{\end{bmatrix}}
\newcommand{\Mpl}{M_{\rm Pl}}

\newcommand{\la}{\lambda}

\newcommand{\ols}[1]{\mskip.5\thinmuskip\overline{\mskip-.5\thinmuskip {#1} \mskip-.5\thinmuskip}\mskip.5\thinmuskip} 
\newcommand{\olsi}[1]{\,\overline{\!{#1}}} 
\makeatletter
\newcommand\closure[1]{
  \tctestifnum{\count@stringtoks{#1}>1} 
  {\ols{#1}} 
  {\olsi{#1}} 
}

\newcommand{\lt}{\left}
\newcommand{\rt}{\right}

\allowdisplaybreaks

\begin{document}

\preprint{YITP-25-93, WUCG-25-07}

\title{Greybody factors of charged 
black holes with axion hair}

\author{Ratchaphat Nakarachinda$^1$}

\author{Petarpa Boonserm$^1$}

\author{Antonio De Felice$^2$}

\author{Shinji Tsujikawa$^3$}

\author{Pitayuth Wongjun$^4$}

\affiliation{
$^1$Department of Mathematics and Computer Science, 
Faculty of Science, 
Chulalongkorn University, Bangkok 10330, Thailand}

\affiliation{
$^2$Center for Gravitational Physics and Quantum Information, Yukawa Institute for Theoretical Physics, Kyoto University, 606-8502, Kyoto, Japan}

\affiliation{
$^3$Department of Physics, Waseda University, 
3-4-1 Okubo, Shinjuku, Tokyo 169-8555, Japan}

\affiliation{
$^4$The Institute for Fundamental Study (IF), 
Naresuan University, 99 Moo 9, 
Tah Poe, Mueang Phitsanulok, Phitsanulok, 65000, Thailand}

\begin{abstract}
We study the greybody factor of charged hairy black holes (BHs) that arise due to the presence of an axion coupled to the electromagnetic field.
Specifically, we consider spin-0 and spin-1 test particles propagating in the background of BHs with axion hair, where the spacetime geometry is modified compared to that of the Reissner-Nordstr\"om (RN) BH. 
In contrast to the RN solution, with a given total BH charge, 
the effective potential for test particles depends on the ratio 
of electric to magnetic charges.
In other words, charged BHs with axion hair breaks 
the electric–magnetic duality present in the RN solution.
We compute the transmission coefficient of test particles 
plunging into the charged hairy BH and find that the deviation from the RN solution is particularly evident for higher multipole moments.
Precise measurements of greybody factors can thus serve as probes 
for the possible existence of axions coupled to the 
electromagnetic field, as well as potential signatures 
of magnetic monopoles.
\end{abstract}

\date{\today}


\maketitle

\section{Introduction}
\label{introsec}

The axion was originally proposed as a solution 
to the strong CP problem in quantum chromodynamics (QCD) 
\cite{Peccei:1977hh,Weinberg:1977ma,Wilczek:1977pj}. 
The original axion model was found to be 
experimentally nonviable, 
leading to its extension through the inclusion of 
heavy quarks \cite{Kim:1979if,Shifman:1979if} or 
an additional Higgs field  \cite{Dine:1981rt,Zhitnitsky:1980tq}. 
In these generalized frameworks, 
the QCD axion can be a good candidate 
for dark matter \cite{Preskill:1982cy,Abbott:1982af,Dine:1982ah} . 
In string theory, antisymmetric form fields naturally give rise to very light axions \cite{Witten:1984dg,Svrcek:2006yi}, 
with masses spanning a wide range, 
$10^{-33}~{\rm eV} \lesssim m_{\phi} 
\lesssim 10^{-10}~{\rm eV}$ \cite{Arvanitaki:2009fg,Acharya:2010zx,Cicoli:2012sz}.
Since such ultralight axions could serve as sources of both dark matter and dark energy, it is of significant interest to investigate their signatures through astrophysical and cosmological observations.

The axion field $\phi$ can interact with photons through 
a Chern-Simons coupling of the form 
$-(\lambda/4) \phi F_{\mu \nu} \tilde{F}^{\mu \nu}$, where $\lambda$ is a coupling constant, 
$F_{\mu \nu}$ is an electromagnetic field strength 
and $\tilde{F}^{\mu \nu}$ is its Hodge dual.
Numerous laboratory and astrophysical observations 
have placed upper bounds on the coupling constant 
$\lambda$ across a wide range of axion masses. 
For the mass range $m_{\phi} \lesssim 10^{-10}~{\rm eV}$, 
the non-observation of gamma rays from the SN1987A event,
converted from axions in the Milky Way's magnetic field,
places a bound of $|\Mpl \lambda| \lesssim 10^7$ \cite{Payez:2014xsa}, 
where $\Mpl=2.4 \times 10^{18}~{\rm GeV}$ 
is the reduced Planck mass. 
The axion with the mass range of order $m_{\phi} \sim 10^{-22}$~eV 
can act as so-called fuzzy dark matter \cite{Hu:2000ke}. 
Based on axion searches involving polarization plane rotation~\cite{Fujita:2018zaj,POLARBEAR:2023ric,Xue:2024zjq}, the current upper bound from the decrease of cosmic microwave background (CMB) polarization is
$|\Mpl \lambda| \lesssim 10^5\,(m_{\phi}/10^{-22}~{\rm eV})$~\cite{Fedderke:2019ajk}.

The presence of axion-photon coupling can give rise to a static, spherically symmetric charged black hole (BH) with axion hair \cite{Lee:1991jw,Visser:1992qh,Garcia:1995qz,Yurova:2002ju,Balakin:2017nbg,Fernandes:2019kmh,Burrage:2023zvk,DeFelice:2024eoj}. This BH solution carries both electric and magnetic charges and features a nontrivial axion profile, distinguishing it from the 
Reissner-Nordstr\"om (RN) BH without scalar hair. 
We note that BHs may acquire magnetic charge through the absorption of 
magnetic monopoles in the early Universe \cite{Stojkovic:2004hz,Kobayashi:2021des,Das:2021wei,Estes:2022buj}. 
If the axion constitutes the main component of dark matter, its clustering may have led to the formation of local structures.
Furthermore, magnetic charge is not neutralized by ordinary matter in conductive media, suggesting that BHs with magnetic charge may be 
more stable than purely 
electric ones \cite{Maldacena:2020skw,Bai:2020ezy}. For a fixed total charge and mass, the quasinormal modes of 
a RN BH are the same, regardless of the ratio 
between electric and magnetic charges \cite{Nomura:2021efi,Pereniguez:2023wxf,
DeFelice:2023rra}. This degeneracy in quasinormal modes, which reflects the electric-magnetic duality, is broken in charged BHs with axion hair \cite{DeFelice:2024eoj}. Consequently, gravitational wave measurements of the ringdown phase in BH merger systems offer a new avenue for distinguishing such hairy BHs from the RN solution.

An intrinsic feature of BHs is their behavior as thermal objects. They are characterized by temperature and entropy and, like blackbody radiators, emit 
the well-known Hawking radiation \cite{Hawking:1974rv,Hawking:1975vcx,Hawking:1976de}.
Indeed, the spectrum of radiation emitted from the BH horizon matches that of an ideal blackbody.
The wave associated with Hawking radiation propagates on a curved spacetime background, and an observer at asymptotic infinity can detect its spectrum.
The spacetime curvature affects the potential barrier experienced by emitted particles, thereby modifying the spectrum of Hawking radiation. The ratio of the observed spectrum at spatial infinity to the spectrum emitted from the BH horizon is known as the greybody 
factor \cite{Page:1976df,Unruh:1976fm}. Mathematically, this phenomenon is equivalent to a scattering problem in quantum mechanics. The greybody factor corresponds to the transmission probability for a particle emitted from a BH--or equivalently, the probability of a particle being absorbed by the BH.

To describe the emission or absorption of spin-2 massless gravitons, the standard approach is to use BH perturbation 
theory on a given spacetime 
background \cite{Berti:2009kk,Pani:2013pma}. 
For the Schwarzschild BH, there are two tensor propagating degrees of freedom, which can be separated into odd- and even-parity sectors. 
The wave function associated with the odd-parity gravitational perturbation satisfies a one-dimensional Schr\"odinger-like equation with an effective potential 
$V_{\rm odd}$, known as the 
Regge-Wheeler potential \cite{Regge:1957td}. 
The even-parity gravitational perturbation also satisfies a similar equation, with a so-called Zerilli
potential $V_{\rm even}$ \cite{Zerilli:1970se}. 
Even though $V_{\rm odd}$ and 
$V_{\rm even}$ differ from each other, they can be expressed in a unified manner using a single 
superpotential \cite{Chandrasekhar:1975nkd}. 
This leads to the equivalence of observables in both the odd- and even-parity sectors-such as the quasinormal modes arising during the ringdown phase of BH binaries \cite{Chandrasekhar:1975zza}.

In the presence of additional fields 
on a strong gravitational background, the analysis of BH perturbations becomes 
more involved. Even for RN BHs, the existence of a vector field leads to a nontrivial mixing between gravitational and electromagnetic 
perturbations \cite{Moncrief:1974gw,Moncrief:1974ng,Moncrief:1975sb,Zerilli:1974ai}. 
To compute the greybody factors and quasinormal modes in such cases, one must solve the coupled differential equations of motion for at least two dynamical fields \cite{Gunter:1980,Kokkotas:1988fm,Leaver:1990zz,DeFelice:2023rra}.
For the charged BH with axion hair discussed above, the axion perturbation is also present, along with two gravitational and two electromagnetic perturbations \cite{Kase:2023kvq}.
In this case, the quasinormal modes associated with the gravitational and electromagnetic perturbations were computed in Ref.~\cite{DeFelice:2024eoj}. 
Still, it is generally nontrivial to unambiguously attribute each quasinormal frequency to its corresponding dynamical degree of freedom.

If we consider test particles other than 
gravitons propagating 
in a static and spherically symmetric 
background, their field equations of motion can be recast into a 
one-dimensional Schr\"odinger-like equation, analogous to the  
Regge-Wheeler and Zerilli equations. 
Here, test particles should be understood to have negligible backreaction on both the spacetime metric and its perturbations. For example, one can consider a scalar particle (spin-0) or a photon (spin-1) plunging into a BH.
In this case, the greybody factor can be computed by analyzing the scattering of test particles through their effective potentials. 

Since the greybody factor cannot generally be obtained analytically, several approximate methods have been developed to compute it.
One such prescription is the WKB 
approximation \cite{Schutz:1985km,Iyer:1986np,Konoplya:2003ii,Matyjasek:2017psv}, which enables the computation of greybody factors and quasinormal modes to a given order of accuracy\footnote{A publicly available code for calculating greybody factors using various orders of the WKB approximation can be found at: \tt{https://goo.gl/nykYGL.}}.
In practice, the WKB method is an approximation that does not yield exact values for the greybody factor.
Instead, the so-called rigorous bound method was developed to derive a lower bound on the greybody factor \cite{Visser:1998ke,Boonserm:2008qf,Boonserm:2009zba} (see also Refs.~\cite{Boonserm:2008zg,Boonserm:2008dk,Boonserm:2009mi,Ngampitipan:2012dq,Boonserm:2013dua,Boonserm:2014rma,Gray:2015xig,Boonserm:2017qcq,Boonserm:2019mon,Barman:2019vst,Chowdhury:2020bdi,Boonserm:2021owk,Javed:2021ymu,Javed:2022pdg,Javed:2023iih,Jha:2023wzo,Heidari:2024bvd,Stashko:2024wuq,Singh:2024nvx,Kanzi:2024ydp}).
An advantage of this method is that the resulting lower bound can be expressed in analytic form. However, the corresponding formula still involves an arbitrary function, which limits its ability to accurately determine the minimum greybody factor of test particles in a given spacetime background.

In this paper, we compute the greybody factor of spin-0 and spin-1 test particles
for charged BHs with axion hair, without 
resorting to the approximations mentioned 
above. For dyonic RN BHs carrying both electric and magnetic charges, the greybody factor depends solely on the total black hole charge and mass. However, the presence of axions coupled to photons breaks this degeneracy, as the transmission coefficient of test particles propagating from infinity to the horizon depends on the ratio between the electric and magnetic charges. 
Moreover, the effective potential varies depending on the axion-photon coupling constant $\lambda$. Therefore, precise investigations of greybody factors may open a new window for probing the existence of magnetic monopoles, as well as axions interacting with photons.
Moreover, it has recently been argued that the greybody factor can serve as a useful tool for modeling the amplitude of gravitational waves during the ringdown phase of BH binaries \cite{Oshita:2022pkc,Oshita:2023cjz,Okabayashi:2024qbz,Rosato:2024arw}. 
While the connection between greybody factors and quasinormal modes has been pointed out for spin-2 gravitons, our study of greybody factors for spin-0 and spin-1 particles may provide a pathway to relate them to quasinormal modes of the corresponding fields.

This paper is organized as follows.
In Sec.~\ref{potential}, we derive the effective potentials for massless spin-0 and spin-1 test particles propagating on a general static and spherically symmetric background.
In Sec.~\ref{potential EMA}, we compute the effective potentials of test particles in Einstein-Maxwell-Axion (EMA) theories and study their dependence on the axion-photon coupling and the ratio between electric and magnetic charges. In Sec.~\ref{photon fall}, we evaluate the lower bound on the greybody factor for spin-1 particles using an analytic formula known in the literature.
In Sec.~\ref{exact vec}, we calculate the greybody factor by directly integrating the wave equation for spin-1 particles 
in EMA theories and analyze the differences in the transmission coefficient compared to the RN BH 
with the same total charge and mass.
In Sec.~\ref{exact sca}, we perform a similar direct integration for the master equation governing spin-0 particles. Sec.~\ref{consec} is devoted to conclusions. The derivation of the lower bound and the discussion on the symmetry of the transmission coefficient are presented in Appendices~\ref{app:low_lim} and \ref{app:mirror_T}, respectively.

\section{Effective potentials of test particles}
\label{potential}

In this section, we derive the effective potentials
governing the motion of test particles in a static, spherically symmetric background.
To this end, we consider 
spin-0 or spin-1 test particles plunging into BHs. 
These particles move under the influence of effective potentials generated in the vicinity of BHs.
The line element of a static and spherically symmetric spacetime is given by
\be
{\rm d}s^2=g_{\mu \nu} {\rm d}x^{\mu}
{\rm d}x^{\nu}=
-f(r) {\rm d}t^2+h^{-1}(r) 
{\rm d}r^2+r^2 \left( {\rm d} \theta^2+
\sin^2 \theta\,{\rm d}\varphi^2 \right)\,, 
\label{metric}
\ee
where $g_{\mu \nu}$ is the metric tensor, 
$f$ and $h$ are 
functions of the areal distance $r$.  
The radius of the event horizon, $r_h$, is defined 
by the conditions $f(r_h)=0=h(r_h)$. 
In the following, we examine how the effective potentials 
for spin-0 (scalar) and spin-1 (photon) particles arise from 
the general line element given in Eq.~(\ref{metric}).

\subsection{Spin-0 test particles}
\label{spin0}

Let us first consider a massless canonical 
scalar field $\chi$ as a test particle. 
It satisfies the Klein-Gordon equation
$g^{\mu \nu} \nabla_{\mu} \nabla_{\nu} \chi=0$, 
where $\nabla_{\mu}$ denotes covariant derivative. 
This equation takes the form 
\be
\partial_{\mu} \left( \sqrt{-g} 
g^{\mu \nu} \partial_{\nu} \chi 
\right)=0\,,
\label{phieq}
\ee
where $\partial_{\mu} \equiv 
\partial/\partial x^{\mu}$, 
and $g$ is the determinant of $g_{\mu \nu}$. 
The spherical symmetry and time independence of 
the line element (\ref{metric}) imply that the scalar 
field can be decomposed as follows:
\be
\chi = \sum_{l, m} e^{-i \omega t} 
\chi_l(r) Y_{lm} (\theta, \varphi)\,,
\label{phiso}
\ee
where $\omega$ is a constant, $\chi_l$ is a function of $r$, 
and $Y_{lm}$ denotes the spherical harmonics that depend 
on $\theta$ and $\varphi$.
We will focus on the $m=0$ mode without loss of generality. 
In this case, $Y_{l0} (\theta)$ satisfies
\be
\frac{1}{\sin \theta} 
\frac{\rd}{\rd \theta} 
\left( \sin \theta \frac{\rd Y_{l0}}
{\rd \theta} \right)
=-l (l+1) Y_{l0}\,.
\label{Yre}
\ee
Substituting Eq.~(\ref{phiso}) into (\ref{phieq}) 
and exploiting the relation (\ref{Yre}), we obtain 
\be
\frac{\sqrt{fh}}{r^2} 
\left( r^2 \sqrt{fh} 
\chi_l' \right)'
+\left[ \omega^2-\frac{l(l+1)f}{r^2} \right] \chi_l=0\,,
\label{scalareq0}
\ee
where a prime denotes the derivative 
with respect to $r$.
We define a rescaled field $u_l$ and a tortoise coordinate $r_*$, as 
\be
u_l(r) \equiv r \chi_l(r)\,,\qquad 
r_* \equiv \int \frac{{\rm d}r}{\sqrt{fh}}\,.
\ee
Then, Eq.~(\ref{scalareq0}) reduces to 
\be
\frac{\rd^2 u_l}{\rd r_*^2} 
+\left[ \omega^2 -V(r) \right]
u_l=0\,,\label{eq:scal_u_schro}
\ee
where 
\be
V(r)=\frac{l(l+1)f}{r^2}
+\frac{\sqrt{fh}}{r} 
\left( \sqrt{fh} \right)'\,.
\label{po0}
\ee
When the background BH solution is not known analytically, 
we can integrate 
Eq.~(\ref{eq:scal_u_schro}) by considering the following asymptotic behaviors of solutions. 
Around the BH horizon, the metric components 
can be expanded as 
\be
f = \sum_{n=1}^{\infty} f_n 
(r-r_h)^n\,,\qquad 
h = \sum_{n=1}^{\infty} 
h_n (r-r_h)^n\,,
\label{fhexpan}
\ee
where $f_n$ and $h_n$ are constants. 
Then, it follows that the potential 
(\ref{po0}) vanishes on the horizon. 

At spatial infinity, we impose the asymptotic 
flatness without a cosmological constant.
Then, the large-distance expansions of 
$f$ and $h$ are given by 
\be
f = 1+\sum_{n=1} \frac{\tilde{f}_n}{r^n}\,,
\qquad 
h = 1+\sum_{n=1} \frac{\tilde{h}_n}{r^n}\,,
\label{fhL}
\ee
where $\tilde{f}_n$ and $\tilde{h}_n$ are constants. This means that $V(r) \to 0$ 
as $r \to \infty$. As long as the potential 
$V(\phi)$ remains positive in the regime $r_h<r<\infty$, it should attain a maximum 
at some intermediate distance.

\subsection{Spin-1 test particles}
\label{spin1}

We now consider the case of a massless photon 
as an example of a spin-1 test particle. 
This is described by a $U(1)$ gauge field 
${\cal A}_{\mu}$, with the corresponding 
Maxwell tensor ${\cal F}_{\mu \nu}=
\partial_{\mu} {\cal A}_{\nu}
-\partial_{\nu} {\cal A}_{\mu}$. 
The field equation of motion for the gauge field is given by\footnote{The following results can also be derived equivalently using the Lagrangian formalism, by assuming that the test field is governed by the action 
${\cal S}_{\rm spin-1}=-(1/4)\int {\rm d}^4 
x\sqrt{-g} {\cal F}_{\mu\nu}
{\cal F}^{\mu\nu}$.} 
\be
\nabla^{\mu} {\cal F}_{\mu \nu}=0\,.
\label{Feq}
\ee
On the background given by (\ref{metric}), 
we choose the configuration of 
${\cal A}_{\mu}$ to be of the form 
\be
{\cal A}_{\mu}=\left( {\cal A}_{t}, 
{\cal A}_{r}, 0, {\cal A}_{\varphi} \right)\,,
\ee
where 
\be
{\cal A}_{t}=\sum_l e^{-i \omega t} 
{\cal A}_{tl}(r) Y_{l 0} (\theta)\,,
\qquad 
{\cal A}_{r}=\sum_l e^{-i \omega t} 
{\cal A}_{rl}(r) Y_{l 0} (\theta)\,,
\qquad 
{\cal A}_{\varphi}=-\sum_l e^{-i \omega t} 
{\cal A}_{\varphi l}(r) 
\sin \theta \frac{\rd Y_{l 0} (\theta)}
{\rd \theta}\,,
\ee
with ${\cal A}_{tl}$, ${\cal A}_{rl}$, 
and ${\cal A}_{\varphi l}$ being functions of $r$. 
The presence of the $U(1)$ gauge symmetry allows 
us to set the $\theta$ component of 
${\cal A}_{\mu}$ zero.
The two fields ${\cal A}_{tl}(r)$ and 
${\cal A}_{rl}(r)$ correspond to those in 
the even-parity sector, while
${\cal A}_{\varphi l}(r)$ arises from 
the odd-parity sector. 

From the $\varphi$ component of 
Eq.~(\ref{Feq}), we obtain 
\be
{\cal A}_{\varphi l}''+\frac{(fh)'}{2fh}
{\cal A}_{\varphi l}'+\frac{1}{fh}
\left[ \omega^2-\frac{l(l+1)f}{r^2} \right]
{\cal A}_{\varphi l}=0\,.
\label{Apl}
\ee
Here, we used relation (\ref{Yre}) and 
its derivative with respect to $\theta$. 
The $\theta$ component of 
Eq.~(\ref{Feq}) can be solved for 
${\cal A}_{tl}$ as
\be
{\cal A}_{tl}=\frac{i}{2\omega} \left[ 
(fh)' {\cal A}_{rl}+2fh{\cal A}_{rl}' \right]\,.
\label{Atl}
\ee
Substituting Eq.~(\ref{Atl}) into the $r$ 
component of Eq.~(\ref{Feq}), we find 
\be
{\cal A}_{rl}''+\frac{3(fh)'}{2fh}{\cal A}_{rl}'
+\frac{1}{fh} \left[ \omega^2-\frac{l(l+1)f}{r^2} 
+f' h'+\frac{1}{2}f'' h+\frac{1}{2}h'' f
\right]{\cal A}_{rl}=0\,.
\label{Arl}
\ee
On the other hand, substituting Eq.~(\ref{Atl}) 
into the $t$-component of Eq.~(\ref{Feq}) 
yields an equation containing the third 
derivative ${\cal A}_{rl}'''$. 
This equation is equivalent to the one obtained by taking the $r$-derivative of Eq.~(\ref{Arl}). 
Thus, we have two independent field equations of motion, Eqs.~(\ref{Apl}) and (\ref{Arl}), which correspond to the two transverse modes of the electromagnetic field.

To cast the equations of motion into a 
Schr\"odinger-like form, we introduce 
the two fields
\be
\psi_1 \equiv (fh)^{1/2}{\cal A}_{rl}\,,\qquad 
\psi_2 \equiv {\cal A}_{\varphi l}\,.
\ee
Then, both $\psi_1$ and $\psi_2$ 
satisfy the same form of equation  
\be
\frac{\rd^2 \psi_i}{\rd r_*^2} 
+\left[ \omega^2 -V(r) \right]
\psi_i=0\,,
\label{eq for spin-1}
\ee
where $i=1,2$, and 
\be
r_*=\int \frac{\rd r}{\sqrt{fh}}
\ee
is the tortoise coordinate. 
The effective potential in 
Eq.~(\ref{eq for spin-1}) 
is given by 
\be
V(r)=\frac{l(l+1)f}{r^2}\,,
\label{po1}
\ee
which is different from that of the spin-0 particle, 
Eq.~(\ref{po0}), in that the latter includes an additional term $(\sqrt{fh}/r)(\sqrt{fh})'$.
Moreover, the potential of the spin-1 test field vanishes for the $l=0$ mode. 
This suggests that the 
black-body radiation corresponding to the $l=0$ 
mode, when observed at asymptotic infinity, 
exhibits the same spectrum as that near the horizon.

\section{Potentials for 
hairy BHs in EMA theories}
\label{potential EMA}

In this section, we revisit the hairy BH solution 
present in EMA theories and compute the effective potentials of spin-0 and 1 test particles 
in this background.
We consider an axion field $\phi$ coupled to an 
electromagnetic field $A_{\mu}$ in the form 
$-(\lambda/4)\phi F_{\mu \nu} \tilde{F}^{\mu \nu}$, 
where $F_{\mu \nu}=\partial_{\mu}A_{\nu}
-\partial_{\nu}A_{\mu}$ 
and $\tilde{F}^{\mu \nu}
=\epsilon^{\mu \nu \rho \sigma} 
F_{\rho \sigma}/(2\sqrt{-g})$ 
with $\epsilon^{0123}=+1$. 
The action of such theories is given by 
\be
{\cal S}=\int{\rm d}^4x\sqrt{-g}
\left[ \frac{\Mpl^2}{2}R-\frac{1}{2}
\partial_\mu \phi
\partial^\mu\phi-\frac{1}{2}m_{\phi}^2 
\phi^2-\frac{1}{4}F_{\mu\nu} F^{\mu\nu}
-\frac{1}{4} \lambda \phi F_{\mu \nu} 
\tilde{F}^{\mu \nu}
\right]\,,
\label{action}
\ee
where $R$ is the Ricci scalar, $m_{\phi}$ is the 
axion mass, and $\lambda$ is the axion-photon coupling constant. We are primarily interested in the case of an ultralight axion with $m_{\phi} \ll r_h^{-1}$, 
where $r_h$ is the horizon radius. For $r_h={\cal O}(10)$~km, this corresponds to 
the mass range $m_{\phi} \ll 
10^{-11}$~eV.
When $m_{\phi} \neq 0$, the background axion field has 
a growing mode $e^{m_{\phi}r}/r$, 
which becomes important at 
the distances 
$r \gtrsim m_{\phi}^{-1}$. 
Since our focus is on the axion 
profile in the regime
$r_h<r \ll m_{\phi}^{-1}$, 
we will set $m_{\phi}=0$ in the following discussion.

\subsection{Consistent 
parameter space of 
hairy BH solutions}

On the background described by Eq.~\eqref{metric}, we consider the gauge field configuration
\be
A_{\mu}=\left[ A_0(r),0,0,-q_M 
\cos\theta \right]\,,
\ee
together with the radial dependent axion field $\phi=\phi(r)$. 
Here, $q_M$ is a constant corresponding to 
a magnetic charge.
The background equations for the fields $f(r), h(r), A_0(r)$ 
and $\phi(r)$ are written, respectively, as follows:
\ba
& &
2\Mpl^2r^3fh'+r^4h A_0'^2+f\lt[q_M^2-2\Mpl^2r^2+h\lt(2\Mpl^2r^2+r^4 \phi'^2\rt)\rt]
=0,\label{e1}\\
& &
2\Mpl^2r^3f'h+r^4h A_0'^2+f\lt[q_M^2-2\Mpl^2r^2+h\lt(2\Mpl^2r^2-r^4\phi'^2 \rt)\rt]
=0\,,\label{e2}\\
& &
\lt(\sqrt{\frac{h}{f}}r^2 A_0'- \la q_M \phi \rt)'= 0\,,\label{e3}\\
& &
\phi''+\lt(\frac{2}{r}+\frac{f'}{2f}+\frac{h'}{2h} \rt)\phi'
-\frac{\la q_M A_0'}{r^2f}\sqrt{\frac{f}{h}}=0\,.
\label{e4}
\ea
Solving Eq.~\eqref{e3} yields
\be
A_0'=\frac{q_E+\la q_M \phi}
{r^2}\sqrt{\frac{f}{h}}\,,
\label{e3 sol}
\ee
where $q_E$ is an integration constant corresponding to the electric charge of BHs.
There are now three independent functions, i.e., 
$f(r)$, $h(r)$, and $\phi(r)$, that should be 
determined by solving Eqs.~\eqref{e1}, \eqref{e2}, and 
\eqref{e4} together with the constraint 
\eqref{e3 sol}. 
On the BH horizon, located at 
$r=r_h$, both $f$ and $h$ vanish.
The boundary conditions around $r=r_h$ given by 
\ba
f &=&
f_1 (r-r_h)+{\cal O}\big[(r-r_h)^2\big],
\label{bc rh1}\\
h &=&
h_1 (r-r_h)+{\cal O}\big[(r-r_h)^2\big]\,, 
\label{bc rh2}\\
\phi &=&
\phi_0+\frac{\la q_M(\la q_M \phi_0+q_E)}
{h_1 r_h^4}(r-r_h)+{\cal O}\big[(r-r_h)^2\big]\,,
\label{bc rh3}
\ea
where $f_1$, $\phi_0$ are constants, and 
\be
h_1=2\Mpl^2r_h^2-q_E^2-q_M^2
-\la q_M \phi_0 \left( \la q_M \phi_0 +2 q_E \right)\,.
\ee
To ensure the property $h(r)>0$ for $r>r_h$, 
we require that $h_1>0$. 
Moreover, we consider the case in which $|\phi(r)|$ is a decreasing function of $r$, 
which leads to the inequality 
$\la q_M(\la q_M \phi_0+q_E)\phi_0<0$. 
A necessary condition for this inequality to hold 
is $\la q_M q_E \phi_0<0$, 
so that 
$q_M \neq 0$ and $q_E \neq 0$. 
In other words, the realization of BHs 
with axion hair requires the presence of both 
nonzero magnetic and electric charges.
Without loss of generality, 
we focus on the case 
\be
q_M>0\,,\qquad q_E>0\,,\qquad 
\phi_0>0\,,\qquad \la<0\,,
\ee
in the following discussion. 
Combining the inequality 
$\la q_M(\la q_M \phi_0+q_E)\phi_0<0$ 
with $h_1>0$, it follows that
\be
\frac{q_E-\sqrt{2\Mpl^2 r_h^2-q_M^2}}
{-q_M \lambda}<\phi_0<
\frac{q_E}{-q_M \lambda}\,,
\label{phi0ra}
\ee
and hence, the allowed range of $\phi_0$ is bounded.

At spatial infinity, we impose asymptotic flatness, 
such that $f \to 1$ and $h \to 1$ as $r \to \infty$. 
We also consider the boundary condition 
$\phi (r \to \infty)=0$ for the axion.
Then, the solutions expanded in the regime 
$r \gg r_h$ are given by 
\ba
f &=& 
1-\frac{M_\text{ADM}}{4\pi\Mpl^2 r}+\frac{q_E^2+q_M^2}{2\Mpl^2r^2}+{\cal O}\left(\frac{1}{r^3}\right)\,,
\label{bc inf1}\\
h &=&
1-\frac{M_\text{ADM}}{4\pi\Mpl^2 r}
+\frac{q_E^2+q_M^2+q_\phi^2}{2\Mpl^2r^2}
+{\cal O}\left(\frac{1}{r^3}\right),\label{bc inf2}\\
\phi &=& 
\frac{q_\phi}{r}+\frac{\lambda q_E q_M}
{2r^2}+{\cal O}\left(\frac{1}{r^3}\right)\,,
\label{bc inf3}
\ea
where $M_\text{ADM}$ and $q_{\phi}$ are constants corresponding to the ADM mass and 
the scalar charge of BHs, respectively.

We are interested in finding hairy BH solutions 
by fixing the ADM mass and total charge as follows:
\be
M\equiv\frac{M_\text{ADM}}{8\pi\Mpl^2}=1,\qquad
q_T\equiv\sqrt{q_E^2+q_M^2}=\sqrt{\frac{13}{50}} \Mpl M\,.
\label{Mq}
\ee
The electric and magnetic  charges are chosen 
to take the forms
\be
q_E=q_T \cos \alpha\,,\qquad 
q_M=q_T \sin \alpha\,,
\ee
respectively, where $\alpha\in[0,\pi/2]$. 
The angle $\alpha=\tan^{-1}(q_M/q_E)$ characterizes 
the ratio between the electric and magnetic charges. 

\begin{figure}[!ht]
\begin{center}
\includegraphics[scale=0.55]{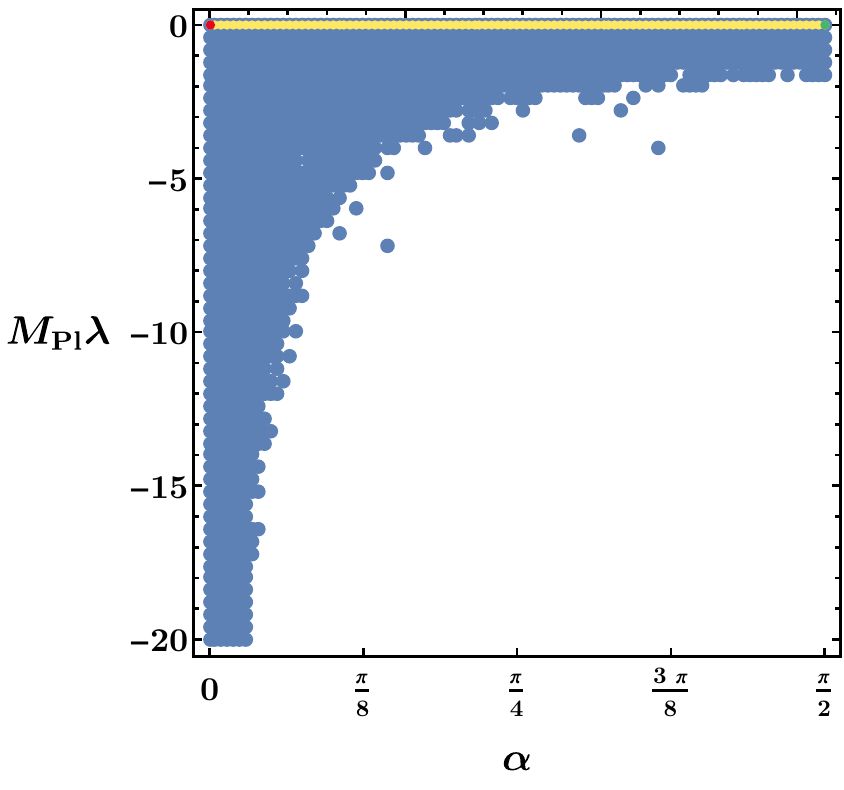}
\end{center}
\vspace{-0.5cm}
\caption{A two-dimensional 
parameter plane
$(\alpha, \Mpl \la)$ showing the existence 
region of hairy BH solutions. 
This corresponds to the case where the ADM mass 
and total charge of BHs are fixed as 
given in Eq.~(\ref{Mq}). 
The field value $\phi_0$ on the horizon 
lies within the range specified by 
Eq.~(\ref{phi0ra}), with 
negative values of $\lambda$.}
\label{fig:region sol}
\end{figure}

For given values of $\alpha$, $\lambda$, 
$f_1$, and $\phi_0$, we can solve the background 
Eqs.~\eqref{e1}-\eqref{e4} by using the 
boundary conditions (\ref{bc rh1})-(\ref{bc rh3})
around $r=r_h$. 
The constants $f_1$ and $\phi_0$ in 
Eqs.~(\ref{bc rh1}) and (\ref{bc rh3}) are 
chosen to satisfy the properties 
$f(r \to \infty)=1$ and $\phi(r \to \infty)=0$ 
at spatial infinity.
The boundary conditions around the horizon are expanded up to fifth order, i.e., 
${\cal O}(r-r_h)^5$, and up to ninth order 
in the region $r \gg r_h$, i.e., 
${\cal O}(1/r^9)$. We numerically integrate 
the background equations of motion outward 
from $r=r_h+10^{-6} M$ to a
sufficiently large distance 
$r=10^2 M$, using the \texttt{NDSolve} 
(with \texttt{Shooting} method option) 
command of \texttt{Mathematica}.

\begin{figure}[!ht]
\begin{center}
\includegraphics[scale=0.58]{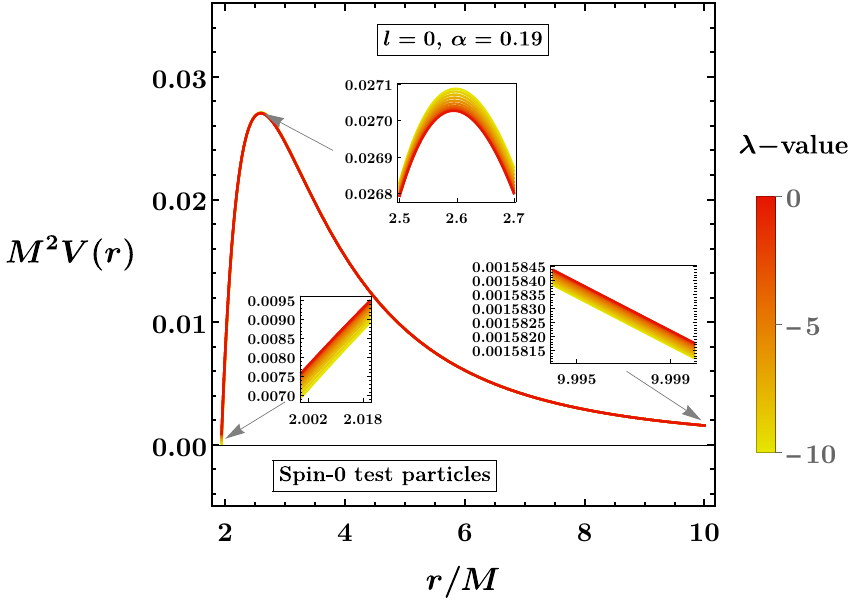}
\hspace{0.5cm}
\includegraphics[scale=0.58]{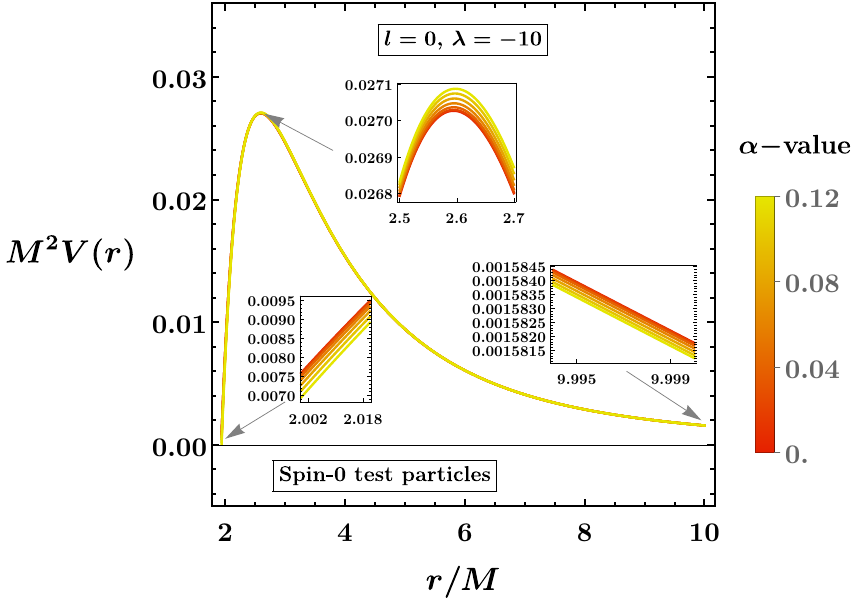}
\\\vspace{0.5cm}
\includegraphics[scale=0.58]{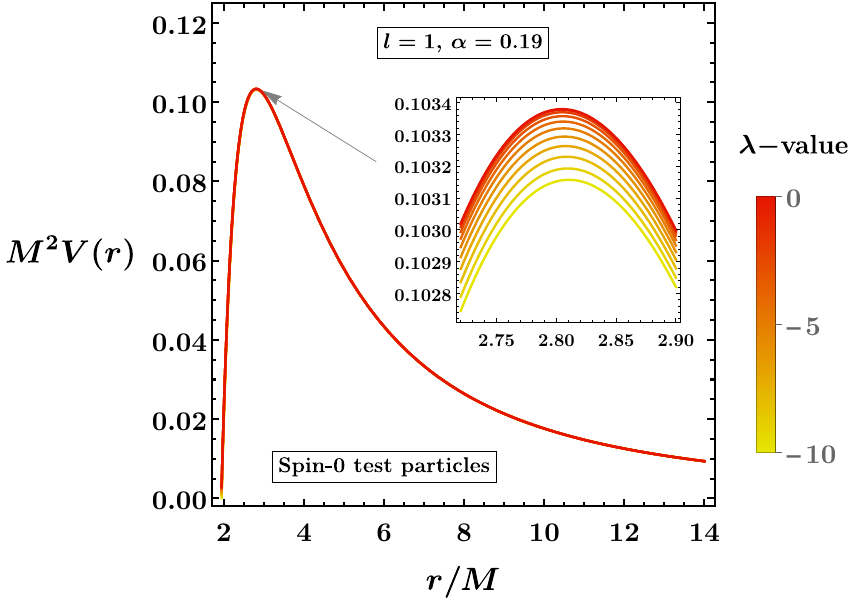}
\hspace{0.5cm}
\includegraphics[scale=0.58]{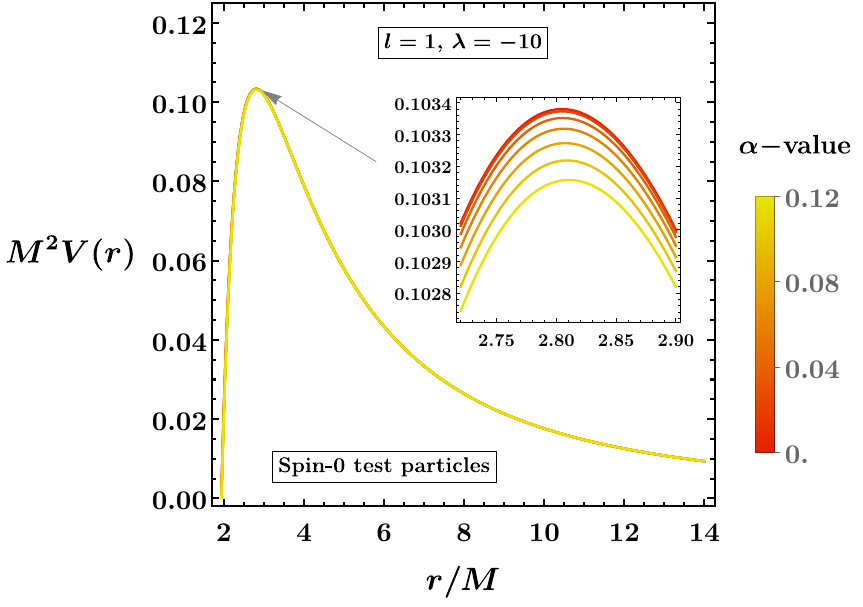}
\end{center}
\vspace{-0.5cm}  
\caption{The effective potentials for spin-0 test particles are shown. 
The top and bottom panels correspond to the $l=0$ 
and $l=1$ modes, respectively. 
The left and right panels illustrate the behavior of the potentials for varying the coupling constant $\lambda$ with fixed $\alpha=0.19$, and for varying the parameter $\alpha$ 
with fixed $\lambda=-10$, respectively.
}
\label{V sca}
\end{figure}

In Fig.~\ref{fig:region sol}, we plot the parameter region where hairy BH solutions exist in the two-dimensional plane $(\alpha, \Mpl \la)$.
Blue dots indicate combinations of parameters that lead to the existence of hairy BH solutions.
For the no-hair BH cases ($\lambda=0$), 
the corresponding parameter sets are represented 
by red, yellow, and green dots.
The red dot at $(\alpha,\la)=(0,0)$ and the green 
dot at $(\alpha,\la)=(\pi/2,0)$ correspond to BHs 
with pure electric charge (standard RN) 
and pure magnetic charge, respectively. The yellow dots represent the solutions with coexisting electric and magnetic charges. As seen in Eq.~(\ref{phi0ra}), the allowed range of $\phi_0$ becomes narrower as $|\lambda|$ 
increases. In Fig.~\ref{fig:region sol}, we 
observe that the amount of magnetic charge 
tends to decrease for larger $|\lambda|$. 
On the other hand, it is possible 
to realize hairy BHs with larger magnetic charges 
by choosing smaller values of $|\lambda|$.
We note that the axion-photon 
coupling in the light mass 
regime ($m_{\phi} \ll 
10^{-11}$~eV) is observationally constrained to be $|\Mpl \lambda| \lesssim 10^5$. The range of $\lambda$ shown in Fig.~\ref{V sca} respects this bound.

To understand the difference 
from the no-hair solution, 
let us consider the RN 
metric given by 
\be
f(r)=h(r)=1-\frac{M_\text{ADM}}{4\pi\Mpl^2 r}
+\frac{q_T^2}{2\Mpl^2r^2}\,,
\ee
where $q_T^2=q_E^2+q_M^2$ is the total squared charge. 
By using this solution, the effective potentials in Eqs.~\eqref{po0} and \eqref{po1} can be expressed 
in terms of the total charge $q_T$.
This implies that the potentials~\eqref{po0} 
and \eqref{po1} remain unchanged as long as 
the total charge is fixed.
It is not possible to distinguish the individual values of the two charges from any signal produced by spin-0 or spin-1 test particles propagating in the background of BHs without axion hair.
As a result, the hairy BH solution with both magnetic and electric charges plays an important role in the search for magnetic monopoles.

%
\subsection{Spin-0 particles}

For a spin-0 test particle plunging into the BH 
with axion hair, we plot the potential given 
by Eq.~(\ref{po0}) for two values of the 
multipole index $l$ in Fig.~\ref{V sca}.
The top and bottom panels show $V(r)$ for $l=0$ and 
$l=1$, respectively, with varying $\lambda$ at 
fixed $\alpha=0.19$ (left) and varying  
$\alpha$ at fixed $\lambda=-10$ (right).
For $l=0$ and $\alpha=0.19$, the peak of 
the corresponding potential 
$V(r)=(\sqrt{fh}/r)(\sqrt{fh})'$ is 
lowest when $\lambda = 0$ (as represented by the red curves).
However, the tails of $V(r)$ for $\lambda=0$ are highest compared to those for the dyonic BH with axion hair.
This implies that the presence of a magnetic charge and/or axionic coupling affects the potential for the $l=0$ mode 
in the following manner:
1) It raises the potential peak, leading to a lower transmission rate of test particles in the high-frequency regime; and
2) it lowers the potential tail, 
resulting in a higher transmission rate in the 
low-frequency regime.
We emphasize that this characteristic arises solely 
from the second term, $(\sqrt{fh}/r)(\sqrt{fh})'$, 
in Eq.~\eqref{po0}. 
For a given value of $\lambda<0$, the peak of the potential increases 
with increasing magnetic charge (i.e., as $\alpha$ increases), while the tail of 
$V(\phi)$ becomes smaller.

For multipole modes with 
$l \geq 1$, the first term 
in Eq.~\eqref{po0}, $l(l+1)f/r^2$, provides the 
dominant contribution to the potential.
Interestingly, this term causes the behavior of $V(r)$ and the corresponding transmission rate to exhibit trends opposite to those caused by the second term in Eq.~\eqref{po0}. 
As shown in the bottom left panel of Fig.~\ref{V sca}, 
for a given value of $\alpha$ in the range 
$0<\alpha<\pi/2$, the potential for $l=1$ tends 
to decrease at all distances as $|\lambda|$ increases. 
This leads to an enhancement 
of the transmission rate 
of test particles.
For a given negative value of $\lambda$, the potential is 
overall suppressed as $\alpha$ increases.
Note that the same trends in the potentials can also be observed for higher multipole modes ($l \geq 2$).

\subsection{Spin-1 particles}
\label{spin1sec}

\begin{figure}[!ht]
\begin{center}
\includegraphics[scale=0.58]{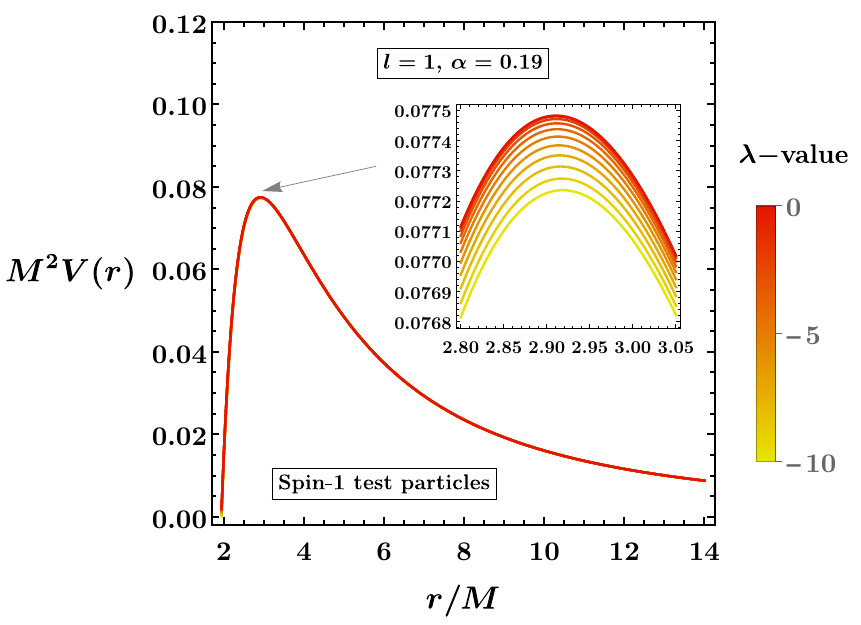}
\hspace{0.5cm}
\includegraphics[scale=0.58]{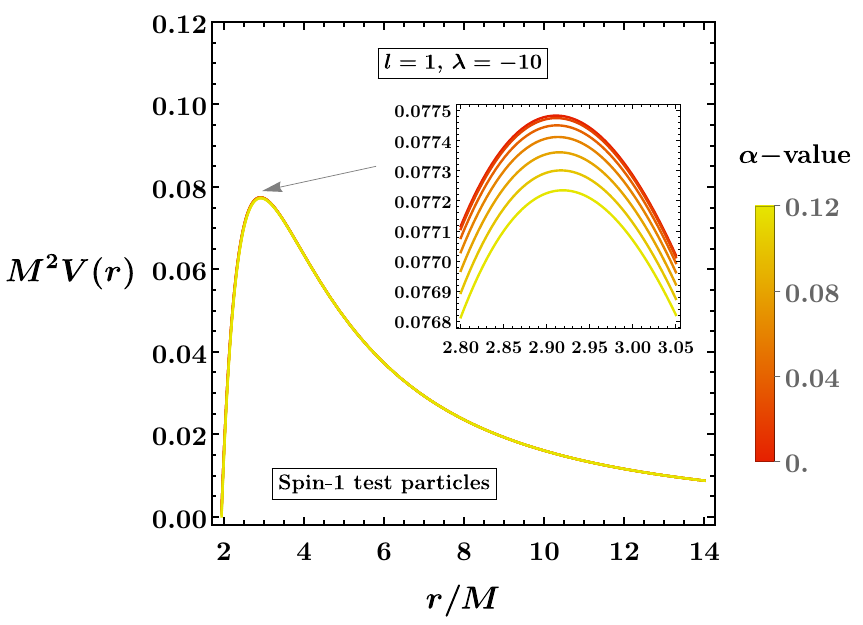}
\end{center}
\vspace{-0.5cm}
\caption{The effective potentials for spin-1 test particles with the $l=1$ mode are shown. 
The left and right panels illustrate the behavior of the potential for varying coupling constant $\lambda$ with fixed $\alpha=0.19$, and for varying magnetic charge (described by the parameter $\alpha$) with fixed $\lambda=-10$, respectively.}
\label{V vec}
\end{figure}

Let us now turn our attention to spin-1 test particles. 
In this case, the potential in Eq.~\eqref{po1} does not include the contribution from 
$(\sqrt{fh}/r)(\sqrt{fh})'$.
This implies that the behavior of $V(r)$ for the spin-1 
case should be similar to that of the spin-0 case with 
$l \geq 1$ modes.
Therefore, the potential is highest when $\lambda=\alpha=0$ 
and decreases in the presence of a magnetic charge and/or axionic coupling. 
These behaviors are illustrated in Fig.~\ref{V vec} 
for the $l=1$ mode.
Thus, the transmission rate of spin-1 particles 
increases with larger values of 
$|\lambda|$ and $|\alpha|$.

\section{Lower bound on the greybody factor 
for a spin-1 partile}
\label{photon fall}

In this section, we study a lower bound on the 
greybody factor by focusing on a spin-1 test 
particle plunging into BHs with axion hair. 
Since the analytic derivation for the lower 
bound on the greybody factor was already performed 
in the literature \cite{Visser:1998ke,Boonserm:2008zg,Boonserm:2009mi}, 
we will quote the main results 
by leaving the detailed derivation of it in 
Appendix \ref{app:low_lim}.
We will show that this bound depends on the magnitude of the magnetic charge and is consistent with the analysis of the test particle potentials presented in Sec.~\ref{potential}.
To obtain the precise values of the greybody factor, we numerically compute it for the spin-1 case in Sec.~\ref{exact vec} and the spin-0 case in 
Sec.~\ref{exact sca}.

As previously shown in Sec.~\ref{spin1}, the two polarization modes of a 
spin-1 test particle, 
denoted by $\psi_1$ and $\psi_2$,
obey the same equation of motion, given in 
Eq.~(\ref{eq for spin-1}).
By collectively denoting both $\psi_1$ and 
$\psi_2$ as $\psi$, the equation is expressed as
\begin{equation}
\psi_{,\mathit{xx}}
+k^{2}(x) \psi=0\,.
\label{psieq}
\end{equation}
where $\psi_{,\mathit{xx}} \equiv 
\partial^2 \psi/\partial x^2$, and   
\be
x \equiv r_*=\int \frac{{\rm d}r}{\sqrt{fh}}\,,
\qquad 
k^2(x) \equiv
\omega^2-\frac{Lf(x)}{r^{2}(x)}
\,,\qquad
L \equiv l(l+1)\,.
\label{xdef}
\ee

By using an arbitrary function $\varphi(x)$, 
the solution to Eq.~(\ref{xdef}) may be written 
in the form \cite{Visser:1998ke,Boonserm:2008qf,Boonserm:2009zba}
\begin{equation}
\psi(x)=a(x)\,\frac{e^{i\varphi}}{\sqrt{\varphi_{,x}}}+b(x)\,\frac{e^{-i\varphi}}{\sqrt{\varphi_{,x}}}\,,
\label{eq:waveab0}
\end{equation}
where $\varphi_{,x} \equiv 
\partial \varphi/\partial x \neq 0$, 
and $a(x)$ and $b(x)$ are functions of $x$.
As we will see in Appendix~\ref{app:low_lim}, 
the conservation of a current $j$ associated with 
the wave function $\psi(x)$ leads to the relation 
$|a(x)|^2-|b(x)|^2=-1$. 
We also find that the transmission coefficient 
$T$ of test particles has the following lower bound:
\be
T \geq T_{\rm min} \equiv \text{sech}^{2} 
\left[\int_{-\infty}^{\infty}\Theta(x)\,{\rm d}x\right]\,,
\label{eq:Tbound0}
\ee
where $\Theta(x)$ is defined in Eq.~(\ref{Theta}). 
Expressing $\Theta(x)$ as a function of $r$, 
we have 
\be
\Theta(r) =\frac{|fh \varphi''(r)
+(fh)'(r)\,\varphi'(r)/2
-i[k^{2}(r)-f h \varphi'(r)^{2}]|}
{2\sqrt{fh}\,\varphi'(r)}\,, 
\ee
so that the integral in Eq.~(\ref{eq:Tbound0}) 
can be expressed as 
$\int_{-\infty}^{\infty}\Theta(x)\,{\rm d}x=\int_{r_{h}}^{\infty} \Theta(r)/\sqrt{fh}\,
{\rm d}r$. For a given function $\varphi(r)$, 
one can compute the minimum transmission 
coefficient, $T_{\rm min}$.
As an example, we choose \cite{Boonserm:2008zg}
\begin{equation}
\varphi_{,x}=\omega,\quad 
{\rm with} \quad 
\omega>0\,,
\label{psichoice}
\end{equation}
which corresponds to 
$\varphi'(r)=\omega/(\sqrt{fh})$. 
In this case, the function $\Theta$ reduces to 
\be
\Theta(r)=\frac{V}{2\omega}
=\frac{fL}{2\omega r^2}\,.
\ee
It then follows that 
\begin{equation}
\int_{-\infty}^{\infty}\Theta(x)\,{\rm d}x
=\frac{L}{2\omega} \int_{r_h}^{\infty} 
\sqrt{\frac{f}{h}}\frac{1}{r^2}
{\rm d}r\,,
\end{equation}
where we have assumed $f>0$, 
since we are only considering
the region outside the horizon. 

In Eq.~(\ref{Mq}), we have defined the quantity 
$M=M_{\rm ADM}/(8 \pi \Mpl^2)$, which has a dimension of length. We then introduce a dimensionless radial coordinate $\bar{r}=r/M$, allowing all quantities 
to be expressed in units of $M$. 
This choice implies setting $M=1$ and requires determining the corresponding value of  
$r_h$ leading to the solution $f=1-2/\bar{r}+\mathcal{O}(\bar{r}^{-2})=h$ in the regime $r\gg r_h$.\footnote{Because the solution is hairy, we generally have $r_h \neq 2$ in these units, due to the non-trivial profile of the axion field.} 
Upon performing this change of coordinates, 
the minimum transmission coefficient is given by
\be
T_{\rm min}=
\text{sech}^{2}\!\left[\frac{L}{2\,\omega M}\int_{\bar{r}_h}^{\infty}\sqrt{\frac{f}{h}}\,
\frac{1}{{\bar r}^2}\,{\rm d}{\bar r}\right]\,,
\label{eq:Tbound}
\ee
where $\bar{r}_h \equiv r_{h}/M$. 

\begin{figure}
\centering
\includegraphics[width=0.48\linewidth]{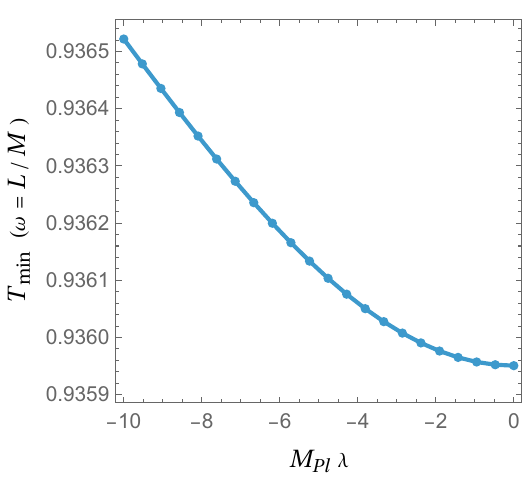}\hfill
\includegraphics[width=0.48\linewidth]{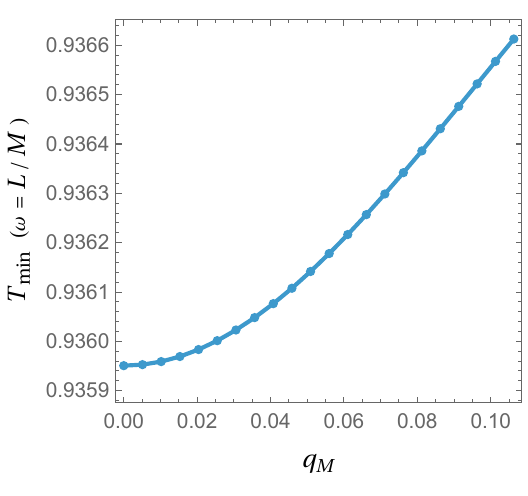}
\caption{In the left panel, we plot the lower bound for $T$ fixing 
$q_{T}=\sqrt{13/2}/5$ and $\alpha=0.19$, with $\lambda$ being varied, and assuming $L/(\omega M)=1$.
On the right, we show instead the behavior of the lower bound for $T$, assuming $L/(\omega M)=1$, but this time changing $q_M$ (or $\alpha$) while keeping $\Mpl\lambda=-10$ fixed together with $q_{T}=\sqrt{13/2}/5$. In the limit $q_M\to0$, we recover 
the RN result. We have fixed the same ADM mass at infinity for all these cases, normalizing $M$ to unity. In the RN case, the solution for the metric components is $f=h=1-2/r+q_T^2/(2r^2)$, leading to $r_h/M=1+\sqrt{87}/10$ and $(2\omega M/L) \,\int_{-\infty}^\infty \Theta\, 
{\rm d}x=M/r_h \simeq 0.5174$.}
\label{fig:fixed_lambda}
\end{figure}

In Fig.~\ref{fig:fixed_lambda}, we plot the minimum 
transmission coefficient $T_{\rm min}$ for 
$\alpha=0.19$ as a function of $\Mpl \lambda$ (left panel) 
and for $\Mpl \lambda=-10$ as a function of 
$q_M=q_T \sin \alpha$ 
(right panel). In both plots, we fix the units by setting the ADM mass of BHs to unity. This implies that, for each value of $\lambda$ (or $\alpha$), a different value of $r_h/M$ 
is obtained. As we discussed in Sec.~\ref{spin1sec}, the potential $V(\phi)=fL/r^2$ tends to be suppressed for nonzero values of $\lambda$ and $\alpha$, 
compared to the case with $\lambda=\alpha=0$. 
In the left panel of Fig.~\ref{fig:fixed_lambda}, we observe that $T_{\rm min}$ slightly increases as $|\lambda|$ increases. 
A similar trend is seen in the right panel, where $T_{\rm min}$, for fixed $\lambda \neq 0$, increases as a function of $\alpha$ (or equivalently, $q_M$). 
These results suggest that the greybody factor can distinguish between BHs with axion hair and RN BHs. 
In the case of hairy 
BHs, the relative ratio of electric to magnetic charge influences the transmission rate of test particles. This breaks the degeneracy present in RN BHs, where $T_{\rm min}$ is fully determined by the ADM mass and total charge, regardless of the mixing angle $\alpha$.

We recall that $\varphi(x)$ is an arbitrary function 
satisfying the condition $\varphi_{,x} \neq 0$.
As a result, the lower bound $T_{\rm min}$ generally depends on the choice of $\varphi (x)$. 
To eliminate the ambiguity inherent in this approach, we will compute the greybody factor numerically in the following sections.

\section{Transmission coefficients for spin-1 test particles}
\label{exact vec}

Up to this point, we have employed a method to 
determine a lower bound for the transmission 
coefficient $T$, based on the choice of an arbitrary function $\varphi(x)$. While this approach is mathematically consistent, different 
choices of 
$\varphi(x)$ can yield different values for the lower bound $T_{\rm min}$. To eliminate this ambiguity and obtain a definitive result, we now proceed with an accurate computation of the greybody factor, without relying on approximations. In this section, we focus on the case of spin-1 test particles, while the analysis for spin-0 test particles will be presented in Sec.~\ref{exact sca}.

We will compute the transmission coefficient through 
a direct approach to solving the wave 
Eq.~\eqref{eq for spin-1} for 
appropriate boundary conditions.
By denoting $\psi_i$ (with $i=1,2$) simply 
as $\psi$ and writing the tortoise coordinate 
as $x=r_*$, Eq.~\eqref{eq for spin-1} 
takes the form 
\be
\psi_{,xx}+\left[ \omega^2
-\frac{Lf}{r^2 (x)}\right]\psi=0\,,
\label{psixeq0}
\ee
where we recall that $L=l(l+1)$. 
In Schwarzschild coordinates, 
Eq.~\eqref{eq for spin-1} can be 
recast in the form
\be
\psi''+\frac{1}{2}\left( \frac{f'}{f}
+\frac{h'}{h} \right) \psi'
+\frac{1}{h} \left( \frac{\omega^2}{f}
-\frac{L}{r^2} \right) \psi=0\,.
\label{eq:chi_eom}
\ee
By using the background equations of motion 
for $f$ and $h$, Eq.~(\ref{eq:chi_eom}) 
takes the form 
\begin{equation}
\psi''={ \frac{2\Mpl^2 r^{2}(h-1)
+\lambda^{2}\phi^{2}q_{M}^{2}+2\lambda q_{E}q_{M}\phi+q_{M}^{2}+q_{E}^{2}}
{2\Mpl^2r^{3} h}\,\psi'}
-\frac{1}{h} 
\left(\frac{\omega^{2}}{f}-\frac{L}{r^{2}}\right) \psi\,.
\end{equation}

In the two asymptotic limits $r\to r_{h}^{+}$ 
and $r\gg r_{h}$, Eq.~(\ref{psixeq0}) reduces to 
\begin{equation}
\psi_{,\mathit{xx}} 
\simeq -\omega^{2} \psi\,,
\end{equation}
whose solution is given by 
\begin{equation}
\psi \propto \exp(\pm i\omega x)\,.
\end{equation}
At this point, we fix the boundary conditions necessary for determining the greybody factor. 
Namely, we impose
\begin{equation}
\psi|_{{\rm horizon}} \simeq 
\frac{1}{\sqrt{\omega}}\,\exp(-i\omega x)\,,\qquad 
\psi|_{{\rm infinity}} \simeq 
\frac{A}{\sqrt{\omega}}\,\exp(-i\omega x)+\frac{B}{\sqrt{\omega}}\,\exp(+i\omega x)\,,
\end{equation}
where $A$ and $B$ are constants.
These boundary conditions describe an infalling photon: part of the incoming wave is reflected by the potential barrier, while the remainder is transmitted into 
the BH. For a given value of $\omega$, we construct a solution to Eq.~(\ref{psixeq0}) that satisfies the boundary conditions specified above.
This construction amounts to determining the constants 
$A$ and $B$, which uniquely characterize the solution. Since the behavior of $\psi$
near the horizon is fully fixed by the purely 
ingoing condition for real $\omega$, 
the constants $A$ and $B$ serve as the only degrees 
of freedom used to reconstruct the solution throughout the horizon exterior.

Let us now reconsider the boundary conditions in more detail. Near the horizon, the metric functions 
$f$ and $h$ can be expanded as in Eq.~(\ref{fhexpan}).
In this region, the tortoise coordinate $x$ is approximately given by
\begin{equation}
x=\frac{1}{\sqrt{f_{1}h_{1}}} 
\ln \left( r-r_{h} \right)\,, 
\end{equation}
where we have set $M=1$ for the normalization 
of the distance.
Then, the wave function near the horizon 
can be expressed as
\be
\psi |_{{\rm horizon}}
=\frac{1}{\sqrt{\omega}}\,(r-r_{h})^{-i\omega/\sqrt{f_{1}h_{1}}}\,.
\ee
Therefore, around the horizon, 
we expand the solution as
\begin{equation}
\psi^{{\rm H}}=\frac{1}{\sqrt{\omega}}\,
(r-r_{h})^{-i\omega/\sqrt{f_{1}h_{1}}}
\sum_{n=0}(\psi^{{\rm H}})^{(n)}\,
(r-r_{h})^{n}\,,
\qquad{\rm with}
\qquad (\psi^{{\rm H}})^{(0)}=1\,.
\label{psiH}
\end{equation}
Here, the coefficients 
$(\psi^{{\rm H}})^{(n)}$ are constants, and the choice $(\psi^{{\rm H}})^{(0)}=1$ 
fixes the overall normalization of the solution.

At spatial infinity, the metric functions 
behave as
\begin{equation}
f \simeq h=1-\frac{2M}{r}+\mathcal{O}(r^{-2})\,,
\end{equation}
so that the tortoise coordinate takes
the form
\begin{equation}
x=r+2M\ln(r-2M)\,.
\end{equation}
Then, the solution satisfying the desired boundary conditions at spatial infinity can be written as
\begin{align}
\psi|_{{\rm infinity}} &
=\frac{A}{\sqrt{\omega}}\,e^{-i\omega r}(r-2M)^{-2i\omega M}+\frac{B}{\sqrt{\omega}}\,e^{i\omega r}(r-2M)^{2i\omega M}\nonumber \\
&\simeq
\frac{A}{\sqrt{\omega}}\,e^{-i\omega r}\,r^{-2i\omega M}+\frac{B}{\sqrt{\omega}}\,e^{i\omega r}\,
r^{2i\omega M}\,,
\end{align}
where the second line holds in the limit $r \to \infty$.
Accordingly, we consider an asymptotic expansion about infinity of the form
\begin{equation}
\psi^{{\rm I}}=\frac{1}{\sqrt{\omega}}\,
e^{-i\omega r}r^{-2i\omega M}\sum_{n=0}\frac{
(\psi_{A}^{{\rm I}})^{(n)}}{r^{n}}+\frac{1}{\sqrt{\omega}}\,e^{i\omega r}r^{2i\omega M}\sum_{n=0}\frac{(\psi_{B}^{{\rm I}})^{(n)}}{r^{n}}\,,
\label{psiI}
\end{equation}
where $(\psi_{A}^{{\rm I}})^{(n)}$ and 
$(\psi_{B}^{{\rm I}})^{(n)}$ are constants.
In this case, the two coefficients $A$ and $B$ can be mapped to 
$(\psi_{A}^{{\rm I}})^{(0)}$ and 
$(\psi_{B}^{{\rm I}})^{(0)}$, respectively.
Once the solution is found, these two coefficients can 
be determined, and the transmission and reflection coefficients are given by
\begin{equation}
T=\frac{1}{|(\psi_{A}^{{\rm I}})^{(0)}|^{2}}\,,\qquad R=\frac{|(\psi_{B}^{{\rm I}})^{(0)}|^{2}}
{|(\psi_{A}^{{\rm I}})^{(0)}|^{2}}\,,
\end{equation}
as shown in Appendix \ref{app:low_lim}.
To ensure consistency, 
one must verify that $T+R=1$.

We now consider the constraints imposed by the equations of motion on the solution in the vicinity of the horizon. 
Near the horizon, the metric functions 
are expanded as Eq.~(\ref{fhexpan}), 
together with the expansion of 
the axion field:
\begin{align}
\phi =\sum_{n=0}
\phi_{n}(r-r_{h})^{n}\,,
\end{align}
where $\phi_{n}$ are constants. 
The wave function $\psi$ is expressed as 
Eq.~(\ref{psiH}) around $r=r_h$.
Then, for instance, the background equations leave the two constants $f_1$ and $\phi_{0}$ as free constants, while all other expansion coefficients are determined as functions of these two.
There are no free constants arising from $\psi$, since we have already fixed 
the normalization by setting
$(\psi^{{\rm H}})^{(0)}=1$.
As a result, all higher-order coefficients 
$(\psi^{{\rm H}})^{(n)}$ ($n>0)$ with 
$n>0$ are uniquely determined by 
$f_1$ and $\phi_0$.
This implies that, once the background solution is determined,\footnote{This method requires first solving the background equations of motion for a chosen set of parameters.} the function $\psi^{\rm H}$ is also fully specified. Consequently, the behavior of 
$\psi$ near the horizon is known.
In particular, we can integrate 
Eq.~(\ref{eq:chi_eom}) by setting the initial conditions for $\psi$ and 
$\psi'$, as $\psi \bigl(r_{h}(1+\epsilon)\bigr)=\psi^{{\rm H}}
\bigl(r_{h}(1+\epsilon)\bigr)$ and $\psi' \bigl(r_{h}(1+\epsilon)\bigr)=
(\psi^{{\rm H}})'\bigl(r_{h}(1+\epsilon)\bigr)$.
We can terminate the integration at an arbitrarily chosen value of $r$, 
for example $r=3M$.

We perform a similar analysis 
in the regime $r\gg r_{h}$. 
In this case, the metric functions are given by Eq.~(\ref{fhL}), with the axion 
field profile:
\be
\phi =\sum_{n=1}\frac{\tilde{\phi}_{n}}{r^{n}}\,,
\ee
where $\tilde{\phi}_{n}$ are constants, 
and $\tilde{h}_1=-2$ in the unit of $M=1$.
At large distances, the wave function takes the form given in Eq.~(\ref{psiI}). 
There are two free constants, $\tilde{f}_{1}$ and $\tilde{\phi}_{1}$, in the expansions of 
$f$, $h$, and $\phi$. All other coefficients in these expansions are determined by 
$\tilde{f}_{1}$ and $\tilde{\phi}_{1}$.
These two constants are related to 
$f_1$ and $\phi_0$, which appear in the 
near-horizon expansions of $f$ and $\phi$.
The values of 
$f_1$ and $\phi_0$ are chosen such that the asymptotic boundary conditions 
$f \to 1$ and $\phi \to 0$ are satisfied at spatial infinity.

For the wave function $\psi^{{\rm I}}$, 
the equation of motion should be regarded as describing two independent solutions. 
That is, we fix the conditions for
$(\psi_{A}^{{\rm I}})^{(n)}$ and $(\psi_{B}^{{\rm I}})^{(n)}$ independently 
of each other. In this case, we find that there are two additional free constants not fixed a priori by the equation of 
motion--namely, $(\psi_{A}^{{\rm I}})^{(0)}$ and $(\psi_{B}^{{\rm I}})^{(0)}$. 
These are precisely the constants 
we aim to determine.
Once the background solution is known, 
they are the only two free constants.
We will fix them as follows. 
By choosing arbitrary values for the two constants, we can construct the solution
$\chi^{{\rm I}}(r)$ without any remaining free parameters. This allows us to impose boundary conditions on $\psi$ at a large 
distance, as 
$\psi \bigl(r_{\infty}\bigr)=
\psi^{{\rm I}}\bigl(r_{\infty}\bigr)$
and $\psi'\bigl(r_{\infty}\bigr)
=(\psi^{{\rm I}})' \bigl(r_{\infty}\bigr)$, where $r_{\infty}$ is chosen such that $r_{\infty}\gg r_{h}$.
We will fix the distance $r_{\infty}$ 
to be $r_{\infty}=100 M$. 

Now, we integrate Eq.~(\ref{eq:chi_eom}) both inward and outward to obtain the values of 
of $\psi(r)$ and $\psi'(r)$ at an intermediate distance $r_m$, 
for example, at $r_m=3M$.
To construct the full solution, we impose the matching conditions at $r=r_m$, by requiring
\be
\psi (r_m^{-})=\psi (r_m^{+})\,,\qquad 
\psi' (r_m^{-})=\psi' (r_m^{+})\,,
\ee
where the left-hand side values come from integrating outward starting near the horizon, and the right-hand side values come from integrating inward starting from 
$r_{\infty}$.
Therefore, we seek the values of
$(\psi_{A}^{{\rm I}})^{(0)}$ and $(\psi_{B}^{{\rm I}})^{(0)}$ that satisfy these two complex matching conditions. 
Once such a solution is found, the scattering problem is well-posed, allowing us to determine both the transmission coefficient
$T$ and the reflection coefficient $R$.
To achieve sufficient numerical precision, we expand the background fields up to order
$n=9$ both near the horizon ($r=r_h$) and in the asymptotic region 
($r \gg r_h$).
For the wave function $\psi$, we use expansions up to order $n=8$ in both regions.

\begin{figure}[ht]
\centering
{\includegraphics[width=0.47\linewidth]{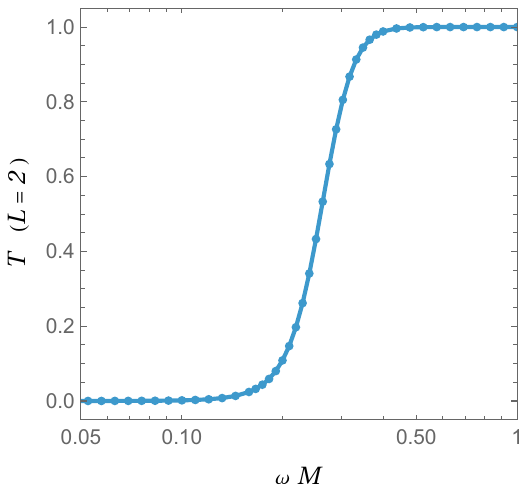}\hfill
\includegraphics[width=0.49\linewidth]{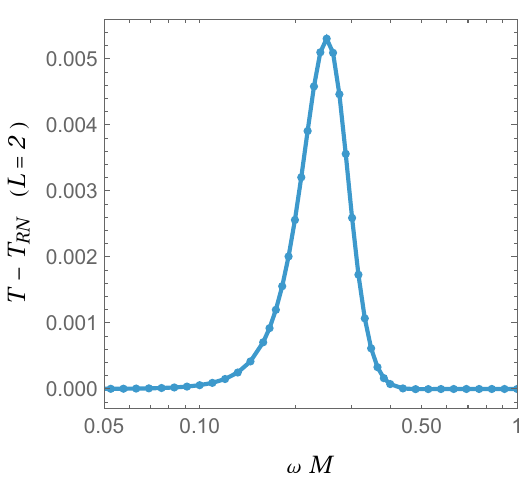}}
\caption{
Left panel: The transmission coefficient as a function of $\omega M$ is shown for the parameters $q_T=\sqrt{13/50}\Mpl M$, $q_M=q_T\sin(1/5)$, $\lambda=-10$, and $l=1$. The BH horizon is located at $r_h \simeq 1.9405$, with the normalization $M=1$. In our numerical calculations, we confirm that $T+R=1$ with an accuracy up to the order of $10^{-8}$. 
Right panel: We plot 
$T-T_{\rm RN}$ as a function of $\omega M$, where $T_{\rm RN}$ corresponds to the transmission coefficient of the RN BH with $q_M=0$ and $\lambda=0$, but with the same ADM mass and total charge.
}
\label{fig:T_exact}
\end{figure}

\begin{figure}[ht]
\centering
{\includegraphics[width=0.47\linewidth]{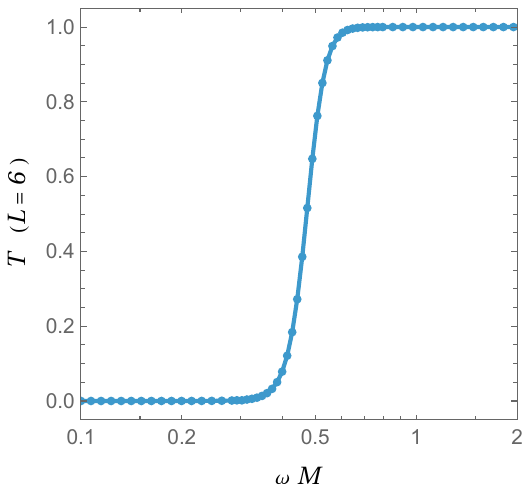}\hfill
\includegraphics[width=0.49\linewidth]{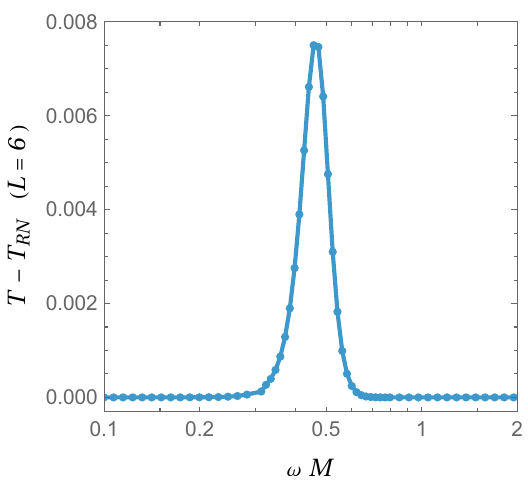}}
\caption{
The transmission coefficient $T$ (left) 
and the difference $T-T_{\rm RN}$ (right)
as a function of $\omega M$, with 
the same values of 
$q_T$, $q_M$, and $\lambda$ as 
in Fig.~\ref{fig:T_exact}, but now with $l=2$.
The position of the horizon remains unchanged.}
\label{fig:T_exact_L6}
\end{figure}

In the left panel of Fig.~\ref{fig:T_exact}, 
we plot the transmission coefficient as a 
function of $\omega M$ for $\alpha=1/5$, 
$\lambda=-10$, and $l=1$ (i.e., $L=2$). 
We observe that low-frequency photons are unable 
to overcome the potential barrier, while 
high-frequency photons can transmit through it. 
The intermediate regime, where $T \simeq 1/2$, 
occurs around $\omega M \simeq 0.25$. 
We recall that in Fig.~\ref{fig:fixed_lambda}, 
we computed $T_{\rm min}$ as a function of 
$q_M$ for the frequency $\omega=l(l+1)/M$.
Since we are now considering the case $l=1$, 
i.e., $\omega M=2$, we find that $T \simeq 1$ 
in the left panel of Fig.~\ref{fig:T_exact}.
This is consistent with the lower bound 
$T_{\rm min} \simeq 0.936$ obtained by 
choosing the function $\varphi$ 
as in Eq.~(\ref{psichoice}).

In the right panel of Fig.~\ref{fig:T_exact}, 
we also plot the difference between $T$ and 
$T_{\rm RN}$, where $T_{\rm RN}$ is computed 
assuming a background with $q_M=0$, 
while keeping the same ADM mass and total charge 
(which corresponds to the RN BH). 
It can be seen that the difference reaches its maximum around $\omega M\simeq 0.25$, with $T>T_{\rm RN}$. 
This implies that, for hairy BHs, more light is transmitted than in the RN case.\footnote{Since 
$0 < T < 1$ holds in any theory, the difference between the transmission coefficients for the hairy BH and RN cases can be at most unity. This limiting case would occur, for example, if $T \approx 1$ for the hairy BH case while $T_{\rm RN} \approx 0$ at the same frequency. However, at sufficiently low or high frequencies, this difference vanishes, as both transmission coefficients approach similar values.} 
This result is consistent with the behavior of the effective potential shown in the right panel of Fig.~\ref{V vec}, where a suppression of $V(\phi)$ is observed for nonzero values of $\alpha$, in contrast to the case $\alpha=0$. 
For large frequencies in the range $\omega M \gtrsim 1$, both $T$ and $T_{\rm RN}$ approach 1; thus, the difference between them tends to zero.

\begin{figure}[ht]
\centering
\includegraphics{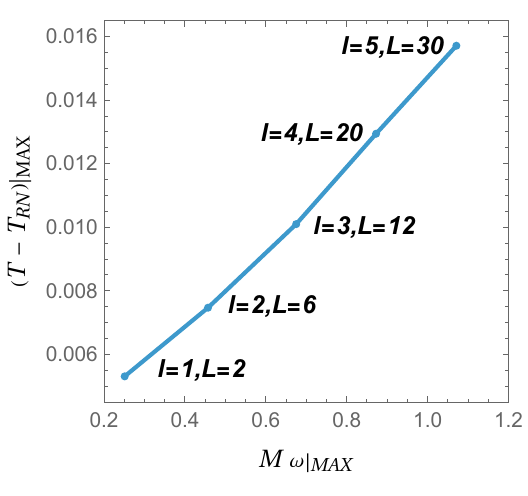}\hfill
\includegraphics{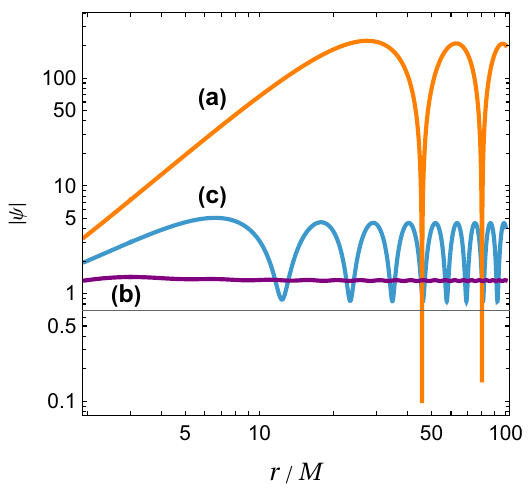}
\caption{Left panel:~Maximum values for $T-T_{\rm RN}$ plotted as a function of the corresponding frequency $\omega$ at which the 
maximum occurs, for five different values of $l$ with $l\in\{1,\dots,5\}$. 
Right panel:~The wave function $|\psi|$ plotted as a function of the radial coordinate $r/M$ for three 
different frequencies: 
(a) $\omega M=9.12\times10^{-2}$ (orange), 
(b) $\omega M=0.575$ (purple), and 
(c) $\omega M=0.263$ (blue). 
The model parameters are chosen to be $q_T=\sqrt{13/50}\Mpl M$, 
$q_M=q_T\sin(1/5)$, $\lambda=-10$, and $l=1$.
In case (a), $|\psi|$ is strongly suppressed near the horizon 
($T \to 0$), whereas in case (b), 
it remains nearly constant across the entire domain ($T \to 1$). 
In case (c), a mild suppression of $|\psi|$ is observed near 
the horizon.
}
\label{fig:T_exact_solutions}
\end{figure}

In Fig.~\ref{fig:T_exact_L6}, 
we also present the behavior of 
$T$ and $T-T_{\rm RN}$, using the same parameter values as in Fig.~\ref{fig:T_exact}, except 
that the multipole index is changed from $l=1$ to $l=2$. 
Although the overall behavior of the transmission coefficient is not drastically different from the case $l=1$, two key differences can be observed: 
1) the transition from low to high values of $T$ occurs at higher frequencies around $\omega M\simeq 0.5$; and 2) the maximum deviation from the RN BH increases to approximately 0.0075, compared to 0.005 for $l=1$.
To study the multipole dependence of the deviation from the RN 
solution, the left panel 
of Fig.~\ref{fig:T_exact_solutions} shows the maximum value of 
$T-T_{\rm RN}$ plotted against the corresponding frequency $\omega$ at which this maximum occurs, 
for various values of $l$ 
in the range $1 \le l \le 5$. 
For these modes, the quantity 
$T-T_{\rm RN}$ increases with 
increasing $l$.
We cannot reliably determine the behavior for large multipoles in the regime $l \gg 1$ due to increasing numerical errors in this regime.
However, the difference 
$T-T_{\rm RN}$ is expected to remain below unity.
In any case, Fig.~\ref{fig:T_exact_solutions} indicates that an observer sending photons into the BH can detect deviations from the RN solution at least at the percent level.

In the right panel of Fig.~\ref{fig:T_exact_solutions}, we show the wave function $|\psi|$ versus $r/M$ for three 
different values of $\omega$. 
The low-frequency mode (case (a)) exhibits strong suppression near the horizon compared to the peak value of $|\psi|$.
The high-frequency mode (case (b)) remains nearly constant outside the horizon.
The intermediate-frequency mode (case (c)) shows mild suppression near the horizon. 
Note that the transmission coefficient quantifies the ratio of the squared amplitudes of the wave before and after traversing the potential barrier.
For instance, in case (a), the amplitude of the oscillations at 
spatial infinity is approximately 200, whereas near the horizon, $|\psi| \simeq 3$.
Note that near the horizon, due to the chosen boundary conditions, we have $\psi\simeq e^{-i\omega x}/\sqrt{\omega}$, so that $|\psi|\to{\rm constant}$. In contrast, at spatial infinity, the different boundary conditions result in oscillatory behavior of $|\psi|$.

Finally, as shown in Appendix~\ref{app:mirror_T}, the transmission coefficient obtained here coincides with that derived for a solution with mirror boundary conditions--that is, for a photon escaping from the BH (e.g., via Hawking radiation) and reaching infinity after traversing the potential barrier.

\section{Transmission coefficients for 
spin-0 test particles}
\label{exact sca}

Let us now compute the transmission coefficient for spin-0 test particles without relying on the approximation employed 
in Sec.~\ref{photon fall}. 
Introducing the notations $\psi=u_l$ 
and $x=r_*$, Eq.~(\ref{eq:scal_u_schro}) 
can be expressed in the form 
\be
\psi_{,xx}+\left[ \omega^2 -V(r) \right] 
\psi=0\,,
\label{psieq2}
\ee
where $V(r)=Lf/r^2+(\sqrt{fh}/r)(\sqrt{fh})'$.
In terms of the Schwarzschild radial 
coordinate $r$, Eq.~(\ref{psieq2}) 
takes the form 
\begin{equation}
\psi''+\frac{1}{2} \left( \frac{f'}{f}
+\frac{h'}{h} \right) \psi'
+\frac{1}{fh} \left[\omega^2
-\frac{2L f+r(f'h+fh')}{2r^2}
\right] \psi=0\,.
\label{psieq2A}
\end{equation}
By using the background equations of motion, Eq.~(\ref{psieq2A}) yields
\ba
\psi''&=&
{\frac{2\Mpl^2 r^{2}(h-1)+\lambda^{2}
\phi^{2}q_{M}^{2}+2\lambda q_{E}q_{M}\phi+q_{M}^{2}+q_{E}^{2}}
{2\Mpl^2r^{3} h}\psi'} \nonumber \\
& &{-\frac{1}{fh}
\left\{ \omega^2-\frac{f[2 r^2 (L - h + 1) \Mpl^2 
- \lambda^2 q_M^2 \phi^2 
- 2 \lambda \phi q_E q_M-q_E^2-q_M^2]}
{2\Mpl^2 r^{4}}
\right\} \psi\,.}
\label{eq:u_sc}
\ea
As long as the axion field $\phi$ remains finite outside the horizon, the potential 
$V(r)$ vanishes in both limits 
$r \to r_h^{+}$ and $r \to \infty$.

\begin{figure}[h]
\centering
\includegraphics[height=3.2in]{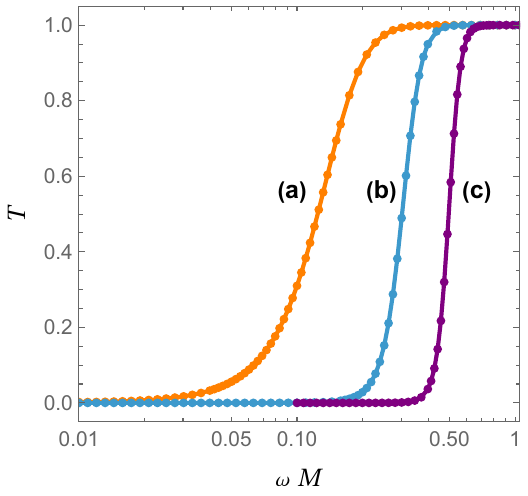}\hfill
\includegraphics[height=3.2in]
{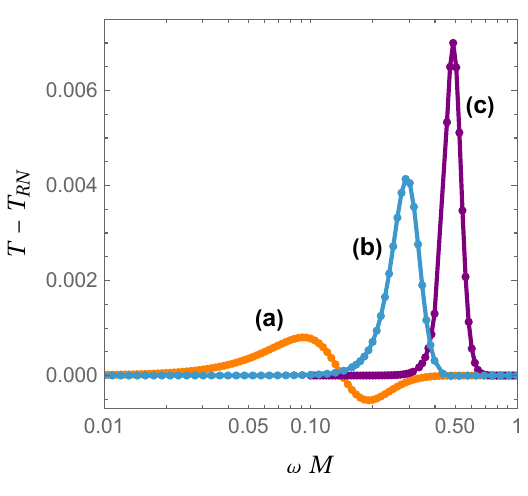}
\caption{Left panel: Transmission coefficient for spin-0 test particles with parameters 
$q_T=\sqrt{13/50}\Mpl M$, $q_M=q_T\sin(1/5)\Mpl$, and 
$\Mpl \lambda=-10$. 
The curves are labeled as (a) 
for $l = 0$, (b) for $l = 1$, 
and (c) for $l = 2$.
Right panel: Plot of $T-T_{\rm RN}$ versus $\omega M$ for $l=0,1,2$, where $T_{\rm RN}$ corresponds to the RN limit with $q_M \to 0$.
The maximum difference $T-T_{\rm RN}$ increases with $l$, and the frequency corresponding to its maximum shifts to higher $\omega$.}
\label{fig:T_exact_L0_scalar}
\end{figure}

As in the case of spin-1 test particles, we seek purely ingoing modes near the horizon, which can be expressed in the following form:
\begin{equation}
\psi|_{\rm horizon}=\frac{1}{\sqrt{\omega}}\,
(r-r_h)^{-i\omega/\sqrt{f_1 h_1}}\sum_{n=0}(\psi^{\rm H})^{(n)}\,(r-r_h)^n\,,
\quad{\rm with} \quad 
(\psi^{\rm H})^{(0)}=1\,,
\end{equation}
whose coefficients $(\psi^{\rm H})^{(n)}$ 
can be determined by Eq.~\eqref{eq:u_sc}. At spatial infinity, we look for a solution of 
the kind
\begin{equation}
\psi|_{\rm infinity}=e^{-i\omega r} r^{-i\omega r_s}\frac{\psi_A(r)}{\sqrt\omega}+e^{i\omega r} r^{i\omega r_s}\frac{\psi_B(r)}{\sqrt\omega}\,,
\end{equation}
where 
\begin{equation}
\psi_A(r)=\sum_{n=0}\frac{(\psi^{\rm I}_A)^{(n)}}{r^n}\,,\qquad
\psi_B(r)=\sum_{n=0}\frac{(\psi^{\rm I}_B)^{(n)}}{r^n}\,.
\end{equation}
As before, Eq.~\eqref{eq:u_sc} imposes constraints on the coefficients of these asymptotic series. 
By integrating from both the horizon and spatial infinity and matching the two solutions, along with their derivatives, at a chosen intermediate point, we can determine the values 
of $(\psi^{\rm I}_A)^{(0)}$ and $(\psi^{\rm I}_B)^{(0)}$. 
The transmission coefficient is then given by 
$T=1/|(\psi^{\rm I}_A)^{(0)}|^2$.

The key difference from the case of spin-1 test particles is the presence of an additional contribution, 
$(\sqrt{fh}/r)(\sqrt{fh})'$, 
to the potential $V(r)$. 
As a result, the potential does not vanish 
even when $l=0$. 
In Fig.~\ref{fig:T_exact_L0_scalar}, we plot the transmission coefficient $T$ for three multipole 
modes (left), along with the difference $T-T_{\rm RN}$
relative to the RN case (right).
For $l=0$, $T$ exceeds 0.5 
for $\omega M>0.1245$.
While $T>T_{\rm RN}$ for 
$\omega M<0.1422$, 
we find $T<T_{\rm RN}$ for 
$\omega M>0.1422$.
As shown in the top panel of Fig.~\ref{V sca}, the peak height of the potential $V(\phi)$ for $l=0$
increases when $\lambda$ and $\alpha$ are nonzero, compared to the case with $\lambda=\alpha=0$.

In the case of $l=1$, the frequency $\omega$ 
required for $T \simeq 1$ is higher than that 
for $l=0$, see Fig.~\ref{fig:T_exact_L0_scalar}. 
Moreover, the difference $T-T_{\rm RN}$ 
remains positive for all values of $\omega$, 
with a peak at 
$\omega M=0.2922$.
These properties arise from 
the fact that, 
for $l \geq 1$, the potential is primarily governed by the term $Lf/r^2$, which dominates over the other contribution $(\sqrt{fh}/r)(\sqrt{fh})'$.
As seen in the bottom panel of 
Fig.~\ref{V sca}, the potential for $l=1$ 
is suppressed when $\lambda$ and $\alpha$ 
are nonzero, compared to the case with 
$\lambda=\alpha=0$.
Thus, the transmission coefficient for $\lambda \neq 0$ and $\alpha \neq 0$ is larger than that of the RN BH.

When $l \geq 2$, the potential barrier becomes higher compared to the case 
of $l=1$. 
As seen in the left panel of Fig.~\ref{fig:T_exact_L0_scalar}, 
the frequency $\omega$ required to reach 
$T \simeq 1$ for $l=2$ is larger than that 
for $l=1$. The right panel also shows that 
the peak value of $T-T_{\rm RN}$ increases 
for $l=2$, compared to the $l=1$. 
In general, the difference between $T$ 
and $T_{\rm RN}$ becomes more pronounced for larger values 
of $l$ of order unity. 
Therefore, the greybody factor provides a useful 
means to distinguish between the hairy BH 
with $\alpha \neq 0$ and 
$\lambda \neq 0$ and the no-hair RN BH with 
$\alpha=0$ and $\lambda=0$.
As demonstrated above, this feature is 
evident for both spin-0 and spin-1 particles.

\section{Conclusions}
\label{consec}

We studied the greybody factors associated with spin-0 and spin-1 test particles plunging into a charged BH with axion hair. 
Compared to the RN BH, the metric components are modified due to the presence of an 
axion-photon coupling of the form $-(\lambda/4) \phi
F_{\mu \nu} \tilde{F}^{\mu \nu}$.
This hairy BH is of the dyonic type, whose existence requires both electric and magnetic charges. The relative contribution of the magnetic charge $q_M$ to the total charge 
$q_T$ is weighed by the 
mixing angle $\alpha$, 
according to 
$q_M=q_T \sin \alpha$.

In Sec.~\ref{potential}, we derived the effective potentials 
$V(r)$ of test particles on a general static and spherically symmetric background described by 
Eq.~(\ref{metric}). 
Both spin-0 and spin-1 test 
particles obey the same form of the Schr\"odinger-type equation, with their effective potentials given by Eqs.~(\ref{po0}) and (\ref{po1}), respectively.
In Sec.~\ref{potential EMA}, 
we first clarified the parameter space in the 
($\alpha, \Mpl \la$) plane 
in which the BH with axion hair 
exists (see Fig.~\ref{fig:region sol}). 
We then examined how the effective potentials depend on the two parameters 
$\lambda$ and $\alpha$. 
As shown in the upper panel of Fig.~\ref{V sca}, the peak of 
$V(r)$ for a spin-0 
particle with $l=0$ becomes higher with increasing values 
of $|\lambda|$ and $\alpha$.
In contrast, the tail regions 
of $V(r)$ exhibit the opposite trend, 
decreasing as $|\lambda|$ 
and $\alpha$ increase.
For spin-0 and spin-1 particles with $l \geq 1$, the effective potentials in the presence of axion hair are overall suppressed compared to those 
of the RN solution.

In Sec.~\ref{photon fall}, we computed the minimum values of the greybody factor using the analytical lower bound given by Eq.~(\ref{eq:Tbound0}). 
The detailed derivation of this formula is provided in Appendix~\ref{app:low_lim}. 
Since the formula involves an arbitrary function $\varphi$, 
the minimum greybody factor 
$T_{\rm min}$ inherits this arbitrariness. We chose $\varphi_{,x}=\omega$ for a 
spin-1 particle and plotted 
$T_{\rm min}$ as functions of
$\Mpl \lambda$ and 
$q_M=q_T \sin \alpha$ in Fig.~\ref{fig:fixed_lambda}.
We found that $T_{\rm min}$
increases with increasing 
values of $|\Mpl \lambda|$ 
and $q_M$, a behavior consistent with the suppression of the effective potential $V$.
Thus, for a fixed total charge 
$q_T$ and ADM mass $M$,
the greybody factor enables us to distinguish between the hairy BH 
($\lambda \neq 0$, $\alpha \neq 0$) and the RN BH ($\lambda=0$, 
$\alpha=0$).

In Sec.~\ref{exact vec}, we carried out a precise calculation of the greybody factors for spin-1 test particles by employing higher-order expansions of the wave functions near the horizon and at spatial infinity.
The transmission rate of spin-1 particles for $\lambda \neq 0$ and $\alpha \neq 0$ 
is greater than that in the RN case, with the difference 
$T-T_{\rm RN}$ reaching its maximum at intermediate frequencies $\omega$ where 
$T \simeq 1/2$. 
The maximum difference 
$T-T_{\rm RN}$ tends to 
increase for larger multipoles 
within the range $l={\cal O}(1)$.
This difference can reach the order of $10^{-2}$, 
and such a deviation may serve as a potential signature for probing charged BHs with axion hair.

In Sec.~\ref{exact sca}, we computed the greybody factors 
for spin-0 particles in the modes $l=0,1,2$. Unlike the case of spin-1 particles, the effective potential for spin-0 particles does not vanish even for $l=0$. In this case, 
we find $T>T_{\rm RN}$ at 
low frequencies $\omega$, 
while $T<T_{\rm RN}$ at high $\omega$. 
For $l \geq 1$, the behavior of 
$T-T_{\rm RN}$ as a function of 
$\omega$ is similar to that observed for spin-1 
particles, in that $T$ is always 
larger than $T_{\rm RN}$. 
The maximum values of 
$T-T_{\rm RN}$, as well as the corresponding frequencies, tend to increase with increasing 
$l$ in the range 
$l={\cal O}(1)$. Thus, the 
measurement of greybody factors 
for spin-0 particles plunging into 
the BH also allows us to probe the 
signature of axions coupled
to photons and the existence 
of magnetic monopoles.

If observational data on greybody factors---alongside quasinormal modes 
of BHs---were to become available, it would offer an additional opportunity to place constraints on the axion-photon coupling constant 
$\lambda$. It would be also of interest to compute the greybody factors for rotating BHs with 
axion hair, e.g., those studied 
in Ref.~\cite{Burrage:2023zvk}.
As an additional benefit, the numerical method employed for the accurate computation of greybody 
factors in this work is broadly applicable and can be extended to BH solutions with nontrivial hair.
We leave such extensions for future investigation.

\section*{Acknowledgements}

ADF and ST thank the members of the Department of Mathematics and Computer Science at Chulalongkorn University for their kind hospitality during their stay, during which this work was initiated.
This research project is supported by the Second Century Fund (C2F), Chulalongkorn University and has received funding support from the NSRF via the Program Management Unit for Human Resources \& Institutional Development,
Research and Innovation [grant number B39G680009].
ST is supported by JSPS KAKENHI Grant Number 22K03642, Waseda University Special Research Projects (Nos.~2025C-488 and 2025R-028), and the FY2025 Grant Program for the Promotion of International Joint Research (No.~2505049) at Waseda University.

\appendix

\section{The lower limit on the transmission 
coefficient}
\label{app:low_lim}

In this Appendix~A, we present the detailed derivation 
of the analytical formula for the lower bound of the transmission coefficient of test particles.
We consider a Schr\"odinger-type equation of the form 
(\ref{psieq}), with the wave function $\psi(x)$, 
namely
\be
\psi_{,xx}+k^2(x) \psi=0\,,
\label{psidiff}
\ee
where $k^2(x)=\omega^2-V(x)$ and $x$ is the tortoise 
coordinate defined in Eq.~(\ref{xdef}). 
We assume that the function $k(x)$ has the following 
asymptotic behavior:
\begin{equation}
k(x) \to k_{\pm}\,,\quad {\rm as} \quad 
x \to \pm \infty\,,
\end{equation}
where $k_{\pm}$ are constants.
Let us now impose the following asymptotic form 
for $x\to+\infty$:
\begin{equation}
\psi_{\infty}=\frac{A}{\sqrt{k_{+}}}
e^{-ik_{+}x}+\frac{B}{\sqrt{k_{+}}}
e^{ik_{+}x}\,,
\label{phiinf}
\end{equation}
which satisfies the equation 
\begin{equation}
\psi_{\infty,xx}=-k_{+}^{2}\,
\psi_{\infty}\,.
\end{equation}
The potential at spatial infinity, denoted by
$V_{+}$, is related to $k_{+}$ through
$k_{+}=\sqrt{\omega^{2}-V_{+}}$.

We now define the current density as
\begin{equation}
j=\frac{1}{2i}\left(\psi^{*} \psi_{,x}
-\psi \psi^{*}_{,x} \right)\,,
\label{jdef}
\end{equation}
where the complex conjugate $\psi^{*}$ satisfies 
the same differential equation as in 
(\ref{psidiff}), namely, 
$\psi^{*}_{,xx}+k^2(x) \psi^{*}=0$.
It then follows that 
\be
j_{,x}=\frac{1}{2i}\left(\psi^{*}\psi_{,xx}
-\psi\psi_{,xx}^{*}\right)=0\,,
\ee
and thus the current density $j$
is conserved everywhere. 
Using the solution (\ref{phiinf}) at 
spatial infinity, we find
\begin{align}
j_{\infty}=\frac{1}{2i}\left(\psi_{\infty}^{*} 
\psi_{\infty,x}-\psi_{\infty} 
\psi^{*}_{\infty,x} \right)
=|B|^{2}-|A|^{2}\,.
\end{align}
Around the BH horizon ($x \to -\infty$), 
we have the following solution 
\begin{equation}
\psi_{-\infty}=\frac{C}{\sqrt{k_{-}}}
e^{-ik_{-}x}+\frac{D}{\sqrt{k_{-}}}
e^{ik_{-}x}\,,
\end{equation}
where $k_{-}$ is a constant. 
The potential around $r=r_h$, which is denoted by 
$V_{-}$, is related to $k_{-}$ as 
$k_{-}=\sqrt{\omega^{2}-V_{-}}$.

Along the same lines, we find that
\begin{equation}
j_{-\infty}=|D|^{2}-|C|^{2}\,.
\label{eq:j_neg_inf}
\end{equation}
Let us discuss the case of a test particle 
propagating from $x \to \infty$ toward 
$x \to -\infty$.
A portion of the wave function is transmitted,
while the remainder is reflected.
In this case, we need to assume that nothing 
propagates from $x \to -\infty$ to the region 
with increasing $x$, so that $D=0$. 
If we impose the normalization condition $C=1$, 
then the conservation of current density implies that 
\be
j_{-\infty} =-1=|B|^{2}-|A|^{2}\,.
\label{eq:j_neg_inf}
\ee
We define the incoming, reflected, and transmitted components of the total wave function as
\begin{align}
\psi_{{\rm inc}} & =\frac{A}{\sqrt{k_{+}}}
e^{-ik_{+}x} \,,\\
\psi_{{\rm ref}} & =\frac{B}{\sqrt{k_{+}}}
e^{ik_{+}x}\,,\\
\psi_{{\rm tra}} & =\psi_{-\infty}(D=0)=\frac{1}{\sqrt{k_{-}}}
e^{-ik_{-}x}\,.
\end{align}
We then define the transmission and 
reflection coefficients as
\begin{equation}
T=\frac{j_{{\rm tra}}}{j_{{\rm inc}}}\,,\qquad 
R=-\frac{j_{{\rm ref}}}{j_{{\rm inc}}}\,,
\end{equation}
where 
\begin{align}
j_{{\rm tra}} & =\frac{1}{2i}\left( \psi_{{\rm tra}}^{*} 
\psi_{{\rm tra},x}-\psi_{{\rm tra}} 
\psi_{{\rm tra},x}^{*}
\right) =-1\,,\\
j_{{\rm ref}} & =\frac{1}{2i}\left( \psi_{{\rm ref}}^{*} \psi_{{\rm ref},x}-\psi_{{\rm ref}} \psi_{{\rm ref},x}^{*}\right)=|B|^{2}\,,\\
j_{{\rm inc}} & =\frac{1}{2i}\left( \psi_{{\rm inc}}^{*}\psi_{{\rm inc},x}
-\psi_{{\rm inc}}\psi_{{\rm inc},x}^{*}\right)
=-|A|^{2}\,.
\end{align}
Therefore, we have
\begin{equation}
T=\frac{1}{|A|^{2}}\,,\qquad 
R=\frac{|B|^{2}}{|A|^{2}}\,,
\end{equation}
and hence
\begin{equation}
T+R=\frac{1}{|A|^{2}}+\frac{|B|^{2}}{|A|^{2}}
=\frac{1+|B|^{2}}{|A|^{2}}=1\,.
\end{equation}

Next, we introduce an arbitrary function 
$\varphi(x)$, such that $\varphi_{,x} \neq 0$
everywhere, and rewrite the general form 
of the wave function as
\begin{equation}
\psi(x)=a(x)\,\frac{e^{i\varphi}}{\sqrt{\varphi_{,x}}}+b(x)\,\frac{e^{-i\varphi}}{\sqrt{\varphi_{,x}}}\,.
\label{eq:waveab}
\end{equation}
Here, we have expressed a general complex 
function in terms of three other complex functions: 
$a,b,\varphi$. Since this representation introduces one redundant degree of freedom, we impose the following condition
\begin{equation}
e^{i\varphi}\,\frac{{\rm d}}{{\rm d}x}\!\left(\frac{a}{\sqrt{\varphi_{,x}}}\right)+e^{-i\varphi}\,
\frac{{\rm d}}{{\rm d}x}\!\left(\frac{b}{\sqrt{\varphi_{,x}}}\right)=0\,,
\end{equation}
which serves as a constraint on $a,b,\varphi$. 
Taking the first and second derivatives of 
$\psi(x)$, it follows that 
\begin{align}
\psi_{,x}  & =(ae^{i\varphi}-be^{-i\varphi})\,
\frac{i\varphi_{,x}}{\sqrt{\varphi_{,x}}}\,,\\
\psi_{,xx} & =-\frac{2i\varphi_{,x} e^{-i\varphi}b_{,x}}{\sqrt{\varphi_{,x}}}-\frac{(\varphi_{,x})^{2}}{\sqrt{\varphi_{,x}}}(ae^{i\varphi}+be^{-i\varphi})+\frac{i\varphi_{,xx}}{\sqrt{\varphi_{,x}}}\,
a e^{i\varphi}\,.
\end{align}
Then, the Schr\"odinger-type equation (\ref{psidiff}) 
can be expressed as 
\begin{equation}
\frac{2i\varphi_{,x}e^{i\varphi}a_{,x}}
{\sqrt{\varphi_{,x}}}-\frac{\varphi_{,x}^{2}}{\sqrt{\varphi_{,x}}}(ae^{i\varphi}+b e^{-i\varphi})-\frac{i\varphi_{,xx}}{\sqrt{\varphi_{,x}}}\,
b e^{-i\varphi}=-k^{2}(x) \left( \frac{ae^{i\varphi}}{\sqrt{\varphi_{,x}}}+\frac{be^{-i\varphi}}{\sqrt{\varphi_{,x}}} \right)\,,
\end{equation}
or as
\begin{equation}
-\frac{2i\varphi_{,x} e^{-i\varphi}b_{,x}}{\sqrt{\varphi_{,x}}}-\frac{\varphi_{,x}^{2}}{\sqrt{\varphi_{,x}}}(ae^{i\varphi}+be^{-i\varphi})+\frac{i\varphi_{,xx}}{\sqrt{\varphi_{,x}}}\,
a e^{i\varphi}=-k^{2}(x)
\left( \frac{ae^{i\varphi}}{\sqrt{\varphi_{,x}}}+\frac{be^{-i\varphi}}{\sqrt{\varphi_{,x}}} \right)\,,
\end{equation}
implying that
\begin{align}
a_{,x} =\frac{1}{2\varphi_{,x}}
\left\{ \varphi_{,xx} 
b e^{-2i\varphi}+i[k^{2}(x)-\varphi_{,x}^{2}](a+be^{-2i\varphi}) \right\}\,,
\end{align}
or
\begin{align}
b_{,x} =\frac{1}{2\varphi_{,x}}
\left\{ \varphi_{,xx} 
ae^{2i\varphi}-i[k^{2}(x)-\varphi_{,x}^{2}](ae^{2i\varphi}+b) \right\}\,.
\end{align}
Since $\varphi$ is arbitrary, 
we may assume that $\varphi$ 
is real and that $\varphi_{,x}>0$.
Then we have
\ba
a^{*}a_{,x}+a a^{*}_{,x}
&=& (|a|^{2})_{,x}
=\frac{1}{\varphi_{,x}}{\rm Re}
\left[ \{ \varphi_{,xx}+i[k^{2}(x)-\varphi_{,x}^{2}]\}\,a^{*}b 
e^{-2i\varphi}\right]\,,\label{are} \\
b^{*}b_{,x}+b b^{*}_{,x}
&=& (|b|^{2})_{,x}
=\frac{1}{\varphi_{,x}}\,{\rm Re}\!
\left[\{\varphi_{,xx}-i[k^{2}(x)-\varphi_{,x}^{2}]\}\,ab^{*}
e^{2i\varphi}\right]\,. \label{bre}
\ea
Using the inequality ${\rm Re}(AB)\leq|A|\,|B|$
for any two complex quantities, 
Eq.~(\ref{are}) leads to 
\be
2|a|\,(|a|)_{,x} \leq
\frac{|\varphi_{,xx}+i[k^{2}(x)-\varphi_{,x}^{2}]|}{\varphi_{,x}}\,|a^{*}be^{-2i\varphi}|\nonumber \\
  =\frac{|\varphi_{,xx}+i[k^{2}(x)-\varphi_{,x}^{2}]|}{\varphi_{,x}}\,|a|\,|b|\,,
\ee
so that 
\begin{equation}
(|a|)_{,x} \leq \frac{|\varphi_{,xx}
+i[k^{2}(x)-\varphi_{,x}^{2}]|}
{2\varphi_{,x}}\,|b|\,.
\end{equation}
Likewise, Eq.~(\ref{bre}) gives
\begin{equation}
(|b|)_{,x}\leq\frac{|\varphi_{,xx}-i[k^{2}(x)-\varphi_{,x}^{2}]|}{2\varphi_{,x}}\,|a|\,.\label{eq:b_x}
\end{equation}

Substituting the ansatz of the wave function, Eq.~\eqref{eq:waveab}, 
into the current (\ref{jdef}), 
we find that 
\be
j=|a|^{2}-|b|^{2}\,.
\ee
By imposing 
\begin{equation}
\lim_{x\to\pm\infty}
\varphi_{,x}
=k_{\pm}\,,
\end{equation}
the normalization at $x \to -\infty$, 
Eq.\ \eqref{eq:j_neg_inf}, gives 
\begin{equation}
|a|^{2}-|b|^{2}=-1\,.   
\end{equation}
Using Eq.~\eqref{eq:b_x}, we obtain
\be
(|b|)_{,x} \leq \frac{|\varphi_{,xx}
-i[k^{2}(x)-\varphi_{,x}^{2}]|}{2\varphi_{,x}}\,
\sqrt{|b|^{2}-1}=\Theta (x)\,
\sqrt{|b|^{2}-1}\,,
\ee
where 
\be
\Theta(x) \equiv 
\frac{|\varphi_{,xx}-i[k^{2}(x)-\varphi_{,x}^{2}]|}{2\varphi_{,x}}=\frac{\sqrt{\varphi_{,xx}^2
+[k^{2}(x)-\varphi_{,x}^{2}]^2}}
{2\varphi_{,x}}\,.
\label{Theta}
\ee
It then follows that 
\be
\int_{|b_{-\infty}|}^{|b_{\infty}|}
\frac{{\rm d}|b|}{\sqrt{|b|^{2}-1}} 
=\left[\text{arccosh}(|b|) \right]_{1}^{|A|} 
\leq \int_{-\infty}^{\infty}\Theta(x)\,{\rm d}x\,,
\ee
or
\begin{equation}
|A|\leq\cosh\!\left[\int_{-\infty}^{\infty}\Theta(x)\,{\rm d}x\right]\,.
\end{equation}
Using this relation, the transmission coefficient 
is bounded as
\begin{equation}
T=\frac{1}{|A|^{2}}\geq \text{sech}^{2}\!\left[\int_{-\infty}^{\infty}\Theta(x)\,{\rm d}x\right]\,.
\label{eq:Tbound}
\end{equation}
Therefore, for a given function $\varphi(x)$, 
we can evaluate the right-hand side of 
Eq.~(\ref{eq:Tbound}), provided that 
the potential $V(x)$ is specified.

\section{Symmetry of the transmission coefficient}\label{app:mirror_T}

In this Appendix~B, we search for a solution to Eq.~(\ref{psidiff}) by integrating outward from the vicinity of the horizon, under the boundary  
conditions $\lim_{x\to\pm\infty}V(x)=0$.
We denote the solution to Eq.~\eqref{psidiff} 
as $\psi_L$, which satisfies the following asymptotic behavior, assuming that 
the parameter $\omega$ is real:
\begin{equation}
\psi_L=\begin{cases}
\dfrac{e^{i\omega x}}{\sqrt\omega}\,,~~~& \text{for } x\to+\infty\,,\\
B_L\dfrac{e^{-i\omega x}}{\sqrt\omega} + A_L\dfrac{e^{i\omega x}}{\sqrt\omega}\,,~~~& \text{for } x\to-\infty\,,
\end{cases}
\end{equation}
where $A_L$ and $B_L$ are constants. 
Then, the transmitted and incident current densities are given by $j_{L,{\rm tra}}=1$ and $j_{L,{\rm inc}}=|A_L|^2$, respectively, so that the transmission coefficient is $T_L=1/|A_L|^2$.

Let us now consider another 
solution to 
Eq.~\eqref{psidiff}, this time obtained by integrating from spatial infinity, as in Appendix~\ref{app:low_lim}.
We denote this solution by $\psi_R$, which satisfies the following boundary conditions:
\begin{equation}
\psi_R=\begin{cases}
A_R\dfrac{e^{-i\omega x}}{\sqrt\omega} + B_R\dfrac{e^{i\omega x}}{\sqrt\omega}\,,~~~& \text{for } x\to+\infty\,,\\
\dfrac{e^{-i\omega x}}{\sqrt\omega}\,,~~~& \text{for } x\to-\infty\,.
\end{cases}
\end{equation}
In this case, the transmitted and incident current densities are
$j_{R,{\rm tra}}=-1$ and 
$j_{R,{\rm inc}}=-|A_R|^2$, respectively, so that the transmission coefficient is given by
$T_R=j_{R,{\rm tra}}/j_{R,{\rm inc}}=1/|A_R|^2$. 
The second solution, $\psi_R$, has been built to have mirror boundary conditions compared to the first solution, $\psi_L$. 
We now show that the two solutions satisfy $T_L=T_R$, indicating that, for this type of Schr\"odinger equation, the transmission coefficient is an intrinsic property of the potential barrier, independent of the direction of incidence.
 
Let us now construct the Wronskian for the two solutions, defined as
\begin{equation}
W=W[\psi_L,\psi_R] \equiv 
\psi_L\psi_{R,x}
-\psi_{L,x} \psi_R\,.
\end{equation}
Taking the $x$ derivative of $W$, 
we obtain
\begin{equation}
W_{,x}=\psi_L \psi_{R,xx}
-\psi_R \psi_{L,xx}=
-k^2(\omega^2,x)\psi_L\psi_R
+k^2(\omega^2,x)
\psi_R \psi_L=0\,.
\end{equation}
This result holds because we have chosen the same frequency for both solutions, i.e., $\omega_L=\omega_R$. As a consequence, the Wronskian $W$ is constant throughout the 
horizon exterior. Evaluating it in the asymptotic regions yields
\begin{align}
\lim_{x\to-\infty}W&=-2iA_L\,,\\
\lim_{x\to+\infty}W&=-2iA_R\,.
\end{align}
The constancy of the Wronskian then implies that $A_L=A_R$, and consequently, $T_L=T_R$.

\bibliographystyle{mybibstyle}
\bibliography{bib2}

\end{document}